\documentclass[final, 5p, longtitle, twocolumn]{elsarticle}
\usepackage{geometry} % see geometry.pdf on how to lay out the page. There's lots.
\usepackage[colorlinks=true,pdfstartview=FitV,linkcolor=blue,citecolor=blue,urlcolor=blue,bookmarks=false]{hyperref}
\usepackage[usenames,dvipsnames]{color}
\usepackage[sc]{mathpazo}
\usepackage[below]{placeins}
\usepackage{afterpage}
\usepackage[separate-uncertainty,retain-explicit-plus,per-mode = symbol, detect-weight=true, range-phrase=\,--\,, range-units=single, tight-spacing=true]{siunitx}
\usepackage{mathtools}
\usepackage{cuted}
\usepackage{longtable}
\usepackage{multirow}
\usepackage[figuresright]{rotating}
\usepackage[mathlines]{lineno}
\usepackage{graphicx}
\usepackage{paralist}
\usepackage{wrapfig}
\usepackage{appendix}
\usepackage{vruler}
\usepackage{etoolbox}
\usepackage{tikz}
\usepackage{listings}
\usepackage{amsmath}
\usepackage{subfigure}
\usepackage{relsize}
\usepackage{etoolbox}
\usepackage{gensymb}
\usepackage{colortbl}
\usepackage{comment}
\usepackage{nicefrac}
\usepackage[version=4]{mhchem}
\usepackage{tabularx}
\newcommand{\todo}[1]{{\color{red}#1}}

\newcommand{\isotope}[2]{\ce{^{#2}#1}}

\DeclareSIUnit\counts{counts}
\DeclareSIUnit\FWHM{FWHM}
\DeclareSIUnit\year{yr}
\DeclareSIUnit\ppt{ppt}
\DeclareSIUnit\ppb{ppb}
\DeclareSIUnit\cal{cal}

\begin{document}
\begin{frontmatter}
\title{Design of a high voltage delivery system \\for noble liquid time projection chambers}
% Define affiliation shortcuts for elsarticle (numbered in order of first appearance)
\newcommand{\PNNL}{1}
\newcommand{\Stanford}{2}
\newcommand{\LLNL}{3}
\newcommand{\Carleton}{4}
\newcommand{\SLAC}{5}
\newcommand{\Bern}{6}
\newcommand{\Alabama}{7}
\newcommand{\SUBATECH}{8}
\newcommand{\UMass}{9}
\newcommand{\Sherbrooke}{10}
\newcommand{\Drexel}{11}
\newcommand{\ITEP}{12}
\newcommand{\UK}{13}
\newcommand{\BNL}{14}
\newcommand{\RPI}{15}
\newcommand{\TRIUMF}{16}
\newcommand{\McGill}{17}
\newcommand{\SNOLAB}{18}
\newcommand{\Laurentian}{19}
\newcommand{\IHEP}{20}
\newcommand{\IME}{21}
\newcommand{\UNCW}{22}
\newcommand{\Yale}{23}
\newcommand{\Windsor}{24}
\newcommand{\ORNL}{25}
\newcommand{\CSU}{26}
\newcommand{\TUM}{27}
\newcommand{\Skyline}{28}
\newcommand{\Mines}{29}
\newcommand{\USD}{30}
\newcommand{\McMaster}{31}
\newcommand{\CUP}{32}
\newcommand{\Montclaire}{33}
\newcommand{\Hawaii}{34}
\newcommand{\UWC}{35}
\newcommand{\UCSD}{36}
\newcommand{\SJTU}{37}
\newcommand{\Erlangen}{38}
\newcommand{\KEK}{39}
\newcommand{\Queens}{40}
\newcommand{\Muenster}{41}
\newcommand{\UBC}{42}      % Not used in current list
\newcommand{\Caltech}{43}  % Not used in current list
\newcommand{\FRIB}{44}     % Not used in current list

%Principal Authors
\author[\PNNL]{R.~Saldanha\corref{cor1}}
\ead{richard.saldanha@pnnl.gov}
\cortext[cor1]{Corresponding authors}
\author[\PNNL]{L.~Pagani\corref{cor1}}
\ead{luca.pagani@pnnl.gov}
\author[\Stanford]{E.~Angelico\fnref{fn1}}
\author[\LLNL]{E.~P.~Bernard}
\author[\Carleton]{B.~Chana\fnref{fn2}}
%\author[\Stanford]{\colorbox{red!30}{J.~Dalmasson}} %No response
\author[\SLAC,\Bern]{S.~Delaquis\fnref{fn3}}
\author[\Stanford]{R.~DeVoe}
\author[\Carleton]{M.~Elbeltagi}
\author[\PNNL]{S.~Ferrara}
\author[\Carleton]{D.~Goeldi\fnref{fn4}}
\author[\Carleton]{R.~Gornea}
%\author[\LLNL]{M.~Heffner} % opted out
% ...

% Footnotes for alternative affiliations and notes
\fntext[fn1]{Now at: TAE Technologies, CA, USA}
\fntext[fn2]{Now at: Canadian Nuclear Laboratories, Chalk River, Ontario K0J 1J0, Canada}
\fntext[fn3]{Deceased}
\fntext[fn4]{Now at: Institute for Particle Physics and Astrophysics, ETH Zürich, Switzerland}
\fntext[fn5]{Also at: University of Chinese Academy of Sciences, Beijing, China}
\fntext[fn6]{Now at: Bates College, Lewiston, ME 04240, USA}
\fntext[fn7]{Now at: Yale University, New Haven, CT 06511, USA}
\fntext[fn8]{Also at: Center for Energy Research and Development, Obafemi Awolowo University, Ile-Ife, 220005 Nigeria}
\fntext[fn9]{Now at: Canon Medical Research USA, Inc.}
\fntext[fn10]{Now at: Research Center for Neutrino Science, Tohoku University, Sendai 980-8578, Japan}

% Address definitions
\address[1]{Pacific Northwest National Laboratory, Richland, WA 99352, USA}
\address[2]{Physics Department, Stanford University, Stanford, CA 94305, USA}
\address[3]{Lawrence Livermore National Laboratory, Livermore, CA 94550, USA}
\address[4]{Department of Physics, Carleton University, Ottawa, ON K1S 5B6, Canada}
\address[5]{SLAC National Accelerator Laboratory, Menlo Park, CA 94025, USA}
\address[6]{LHEP, Albert Einstein Center, University of Bern, 3012 Bern, Switzerland}
\address[7]{Department of Physics and Astronomy, University of Alabama, Tuscaloosa, AL 35405, USA}
\address[8]{SUBATECH, Nantes Universit\'e, IMT Atlantique, CNRS\/IN2P3, Nantes 44307, France}
\address[9]{Amherst Center for Fundamental Interactions and Physics Department, University of Massachusetts, Amherst, MA 01003, USA}
\address[10]{Universit\'e de Sherbrooke, Sherbrooke, QC J1K 2R1, Canada}
\address[11]{Department of Physics, Drexel University, Philadelphia, PA 19104, USA}
\address[12]{National Research Center ``Kurchatov Institute'', Moscow, 123182, Russia}
\address[13]{Department of Physics and Astronomy, University of Kentucky, Lexington, KY 40506, USA}
\address[14]{Brookhaven National Laboratory, Upton, NY 11973, USA}
\address[15]{Department of Physics, Applied Physics, and Astronomy, Rensselaer Polytechnic Institute, Troy, NY 12180, USA}
\address[16]{TRIUMF, Vancouver, BC V6T 2A3, Canada}
\address[17]{Physics Department, McGill University, Montr\'eal, QC H3A 2T8, Canada}
\address[18]{SNOLAB, Lively, ON P3Y 1N2, Canada}
\address[19]{School of Natural Sciences, Laurentian University, Sudbury, ON P3E 2C6, Canada}
\address[20]{Institute of High Energy Physics, Chinese Academy of Sciences, Beijing, 100049, China}
\address[21]{Institute of Microelectronics, Chinese Academy of Sciences, Beijing, 100029, China}
\address[22]{Department of Physics and Physical Oceanography, University of North Carolina Wilmington, Wilmington, NC 28403, USA}
\address[23]{Wright Laboratory, Department of Physics, Yale University, New Haven, CT 06511, USA}
\address[24]{Department of Physics, University of Windsor, Windsor, ON N9B 3P4, Canada}
\address[25]{Oak Ridge National Laboratory, Oak Ridge, TN 37831, USA}
\address[26]{Physics Department, Colorado State University, Fort Collins, CO 80523, USA}
\address[27]{Physikdepartment and Excellence Cluster Universe, Technische Universit{\"a}t M{\"u}nchen, Garching 80805, Germany}
\address[28]{Skyline College, San Bruno, CA 94066, USA}
\address[29]{Department of Physics, Colorado School of Mines, Golden, CO 80401, USA}
\address[30]{Department of Physics, University of South Dakota, Vermillion, SD 57069, USA}
\address[31]{Department of Physics and Astronomy, McMaster University, Hamilton, ON L8S 4M1, Canada}
\address[32]{IBS Center for Underground Physics, Daejeon, 34126, South Korea}
\address[33]{Department of Physics and Astronomy, Montclair State University, Montclair, NJ 07043, USA}
\address[34]{Department of Physics and Astronomy, University of Hawaii at Manoa, Honolulu, HI 96822, USA}
\address[35]{Department of Physics and Astronomy, University of the Western Cape, P\/B X17 Bellville 7535, South Africa}
\address[36]{Physics Department, University of California San Diego, La Jolla, CA 92093, USA}
\address[37]{School of Physics and Astronomy, Shanghai Jiao Tong University, Shanghai 200240, China}
\address[38]{Erlangen Centre for Astroparticle Physics (ECAP), Friedrich-Alexander University Erlangen-N{\"u}rnberg, Erlangen 91058, Germany}
\address[39]{KEK, High Energy Accelerator Research Organization 1-1 Oho, Tsukuba, Ibaraki 305-0801, Japan}
\address[40]{Department of Physics, Queen's University, Kingston, ON K7L 3N6, Canada}
\address[41]{Institut f{\"u}r Kernphysik, Westf{\"a}lische Wilhelms-Universit{\"a}t M{\"u}nster, M{\"u}nster 48149, Germany}
\date{} % delete this line to display the current date

%\linenumbers
\begin{abstract}
Noble liquid time projection chambers (TPCs) are a leading technology in the detection of ionizing radiation, particularly in applications such as accelerator neutrino physics, dark matter detection, and neutrinoless double beta decay. This paper addresses the design considerations for implementing stable high voltage (HV) systems within large noble liquid TPCs, with a focus on the nEXO experiment. Utilizing insights from prior HV research and experimental investigations, we outline factors influencing HV stability and discuss design choices to improve stability and prevent electrical discharges. A novel HV delivery system concept is presented, tailored for the nEXO TPC, which incorporates these design considerations while also meeting the stringent radiopurity requirements of the nEXO neutrinoless double beta decay search. These design considerations and their specific implementation towards a HV delivery system offer guidance to future experiments applying high voltage in noble liquid environments.
\end{abstract}
\end{frontmatter}

\section{Introduction}
Noble liquid time projection chambers (TPCs) \cite{nygren1974proposal, chen1976neutrino, rubbia1977liquid} are one of the leading detector technologies for the detection of ionizing radiation and have been widely used in the fields of accelerator neutrino physics \cite{acciarri2017design, abi2020dune}, direct detection of dark matter \cite{benetti2008warp, agnes2015darkside, akerib2019lz, aprile2024xenon, meng2021pandax}, and the search for neutrinoless double beta decay \cite{exo2012exo200, adhikari2022nexo}. The widespread use of TPCs is in part due to their ability to accurately reconstruct the energy, position, and tracks of particle interactions in large sensitive volumes. These capabilities are enabled by the application of an electric field, typically \SIrange{100}{500}{\volt\per\cm}, to uniformly drift ionization charge towards a readout plane at one end of the active volume. As experiments push to higher sensitivities, TPC detectors with larger active volumes are required, with drift lengths of current and future generations of experiments ranging between \SIrange{1}{6}{\meter}. Maintaining the required electric field requires voltage differences of \SIrange{10}{300}{\kilo\volt} between the electrodes that define the TPC drift volume.

The application of stable high voltage (HV) in noble liquid detectors has historically proven difficult \cite{rebel2014high}. The need to limit electronegative contaminants to maintain long electron drift lengths and the stringent radiopurity constraints for rare event searches severely limit the amount and type of insulating material that can be deployed. Additionally, in some experiments the cost of the noble liquid (e.g., isotopically enriched xenon or argon derived from underground sources) or specific background interaction topologies \cite{lux2017signal}, constrain the size of the inactive volume between the HV electrodes and the surrounding grounded components. This often leads to enhanced fields outside the TPC drift region. Designing a TPC to operate stably at high voltage for long periods of time with all of these constraints is challenging and many experiments have had to operate below their target drift electric field due to high voltage instabilities producing light and charge inside the detector.
These difficulties have led to numerous studies of HV behavior in noble liquids. We note that in addition to studies within the field of particle physics, there is a vast amount of applicable literature and knowledge from the field of HV engineering and its applications to electrical power transmission and HV devices~\cite{latham1995high, kuffel2000high, kuchler2017high}. In this paper, we discuss the design considerations that are important to TPC HV systems. We draw from a wide range of HV studies across several disciplines, and present a design concept for a HV delivery system for the nEXO neutrinoless double beta decay experiment~\cite{adhikari2022nexo} along with recommendations that can be used to design similar systems. 

\section{Design considerations for long term HV stability}
\label{sec:hvdesign}
The dielectric strengths of liquid helium, argon, and xenon have been measured in small-scale, controlled setups to be in excess of \SI{100}{kV/cm}~\cite{swan1961influence, gerhold1998properties,rebel2014high}. However, as noted above, nearly all large noble liquid TPCs have observed the onset of instabilities and even discharges at average fields far below this value~\cite{rebel2014high}. 
There have been several experimental investigations into the causes of these issues, and while there is still no fundamental understanding at the micro-physics level~\cite{sun2016formation, lesaint2016prebreakdown}, several contributing factors have been identified. In this section we discuss some key aspects that should be taken into account in the design of noble liquid HV systems and suggest HV engineering techniques to improve long-term stability at high voltage.

\subsection{Maximum electric field strength}
The maximum electric field strength in the noble liquid is one of the primary considerations for long term stability at high voltages. Based on empirical results in test setups, several experiments aim to limit the maximum electric field in the noble liquid\footnote{In some cases the limits are applied only to cathodic surfaces~\cite{LZTDR}.} to less than \SIrange{40}{50}{kV/cm}~\cite{LZTDR, nexo2018sensitivity, aalseth2018darkside}, though as we discuss later in Sec.~\ref{sec:area_effects}, this value should also depend on the size of the electrodes. While the average electric field between electrodes can be easily calculated by the ratio of the voltage difference to the electrode spacing, there are several factors that can cause localized field enhancements. The strength of the electric field is enhanced by curvature of conducting surfaces~\cite{bhattacharya2016dependence} with significant increases in the electric field at the surfaces of wires, edges of thin conductive coatings, and other sharp conductive edges (screw heads, threads, etc.). Analytical calculations of the field enhancement are often not possible and detailed Finite Element Analysis (FEA) modeling is required to ensure that the maximum fields do not exceed the design requirements. It should be noted that because the electric field depends on the curvature of the surface and components such as HV connections and wires often break the axial symmetry of the TPC detector, a full three dimensional FEA model is often required to obtain accurate estimates of the electric field. 

A typical mitigating design strategy is to decrease the curvature of the conductors as much as allowed by other constraints and to shield any sharp edges with smooth conducting surfaces. Additionally, strong field enhancements can occur at ``triple points'' --- junctions where three different materials (e.g. conductor-insulator-xenon) meet. Due to the different dielectric constants at the junction, the electric field is typically very large and triple points are often the source of electron emission that can initiate a secondary emission avalanche~\cite{jordan2007electric}. The field at the triple point can be reduced by recessing, metalizing, or shielding the junction where possible~\cite{wetzer1995hv,latham1995high, faircloth2014technological}. It should be noted that the electric field at a mathematically ideal triple point can be infinite~\cite{hurd1976edge, jordan2007electric} and hence when using FEA models, the calculated maximum electric field at the junction strongly depends on the mesh refinement used, making it difficult to obtain reliable estimates. Nevertheless, methods using a volume-averaged electric field around the triple point still allow for meaningful comparisons between different FEA model geometries and materials~\cite{beneventi2018new}.

\subsection{Area and volume effects}
\label{sec:area_effects}

As next generation noble liquid TPCs push to larger active volumes, the size of the electrodes needed to apply the desired electric field continues to increase. 
Full-scale tests of electrodes in cryogenic liquids are expensive and time-consuming; an understanding of the scaling factors that allow for stable full-scale operation through extrapolation from smaller tests is very valuable. It has been well established for a wide range of dielectric media that the electric field at which breakdown occurs decreases as the electrode area or inter-electrode volume increases~\cite{weber1956area,mazurek1987energy}. Over the last few decades this inverse relationship (scaling effect) has also been demonstrated in noble liquids such as helium~\cite{okubo1996high, phan2021study}, argon~\cite{acciarri2014liquid,auger2016electric}, and xenon~\cite{tvrznikova2019direct, watson2023study}. 

There are two main hypotheses for this scaling effect. One is a statistical argument based on the extreme value or weakest link theory~\cite{cross1982physical} where high voltage instabilities are presumed to be initiated by random independently distributed features (e.g. surface irregularities on electrodes that cause localized field enhancements). 
The larger the area or volume, the larger the probability of having a feature that initiates a breakdown. The second argument is related to locally stored energy, where the larger volume of electric field stores more electrostatic energy, which is then available for the propagation of a discharge~\cite{mazurek1987energy}. While a quantitative physical model of this effect does not exist yet, experimental measurements of the scaling of the breakdown field with area are now available in the literature for liquid helium~\cite{phan2021study}, argon~\cite{auger2016electric}, and xenon~\cite{watson2023study}, though there is considerable scatter in the points from different experimental setups and there is some debate about the functional form to use for the extrapolation~\cite{phan2021study}.

\subsection{Materials}
The choice of materials for rare-event detector components is normally strongly constrained by requirements on radiopurity, outgassing, and chemical purity of the noble liquid. Nevertheless, there are some generally accepted guidelines that should be followed when possible.

\subsubsection{Electrodes}
It is known that the choice of electrode material, particularly for the cathode, can affect the breakdown voltage. Empirical studies have found that electrodes made from materials like stainless steel and titanium are considerably more stable than electrodes made from materials like aluminum and copper~\cite{descoeudres2009dc,latham1995high}. This is believed to be related to the work function, material hardness, resistance to corrosion, and the presence of insulating or semiconducting oxide layers on the surface~\cite{tomas2018study}.

High voltage breakdown is often initiated by local electric field enhancements at defects or sharp asperities on the electrode surface, and hence the surface finish and condition of electrodes plays an important role~\cite{okubo1996high, phan2021study}. Standard practice includes polishing electrodes using either mechanical, chemical, or electrochemical processes~\cite{latham1995high}.
 
Mechanical polishing is most common, though extreme care must be taken to remove all traces of polishing material from the surface. Microscopic, dielectric abrasives from mechanical polishing can become embedded in the surfaces of metal electrodes, and cannot be removed by ultrasonic or other mechanical cleaning methods~\cite{hryhorenko2023, Viklund2024}. Insulating particles on the surfaces of electrodes act as local field enhancing asperities with similar enhancement strength as conducting asperities, depending on their relative dielectric constant~\cite{latham1995high, kuffel2000high}. These particles can support hopping conduction and charged surface states that promote the injection of electrons initiating high voltage phenomena~\cite{latham1995high}. Finally, residual polishing compounds can add unwanted radioactivity.
 
Chemical and electrochemical polishing techniques, which can alter the oxide layer and the chemical composition of the electrodes at the surface, have also been shown to be effective at reducing HV instability ~\cite{tomas2018study}. These techniques may be preferable to mechanical polishing techniques, as they avoid exposure to dielectric abrasives which may embed into electrode surfaces.
 
To improve HV stability, the coating of electrodes with \SI{100}{\nano\metre}-scale films of a variety of metal and insulating materials has been explored~\cite{jedynak1964, StanfordHV, lincpad2025}. There is growing evidence that HV phenomena may be mitigated by coating electrodes with specially selected materials such as platinum, \ce{MgF2}, \ce{SiO}, and polymers. One effect, which is independent of the chemical properties of the deposited material (such as work function or hardness), may be that these coatings reduce the field enhancement of sub-micron embedded dielectric abrasives or other metallic surface asperities. When considering conductive coatings on detector components, care should be taken to ensure that particulate from the coating is not inadvertently released during assembly and installation. Particulate in the noble liquid dielectric can cause enhanced fields and lead to HV instability and discharges~\cite{pan2020review, kuffel2000high}. Radioactivity measurements of some plating materials can be found in Ref.~\cite{Leonard2017radanalysis}. 

\subsubsection{Insulators}
\label{sec:insulators}
Insulators are needed to mechanically support and electrically insulate HV electrodes. The bulk dielectric strength of most solid insulators is significantly higher than that of noble liquids; hence insulators are also used to displace the noble liquid from regions of high field, though radiopurity and outgassing considerations often limit the amount of solid insulation used. However the surface of the insulator is often vulnerable to surface flashovers, especially in the presence of a triple point at the connection with the electrode. Studies have found that insulating materials with higher homogeneity and lower relative permittivity serve as better insulators~\cite{latham1995high}. Materials with higher permittivity than the noble liquid can enhance the electric field in the liquid, increasing the likelihood of a flashover.

\subsection{Insulator surface geometry}
Most models of surface flashover involve electron cascades that initiate at a triple point near the cathode and travel across the surface of the insulator~\cite{latham1995high}. There are several HV engineering techniques that can be used to reduce the likelihood of a surface flashover. The probability of producing secondary electron avalanches can be reduced by carefully selecting the angle of the junction between the insulator and electrode~\cite{wetzer1995hv, jordan2007electric} such that initial seed electrons are less likely to impinge on the insulator surface. Another key parameter is creepage length - the shortest distance between two electrodes (at different voltages) along the surface of an insulator. The required creepage length depends on many factors, but it is generally recommended to keep the field along the surface $<$\SIrange{1}{5}{\kilo\volt\per\centi\metre}~\cite{battel2012high, faircloth2014technological}. Ribs are typically cut into the insulator surface, which not only increase the creepage length, but also introduce regions where the surface is perpendicular to the field, trapping charge on the surface and also possibly reducing the field at the triple point~\cite{yamamoto1996effects, yamamoto1997numerical}. It should be noted that while increasing creepage length and adding ribbing can help, several studies indicate that the most critical parameter for reducing the likelihood of surface flashover is to reduce the electric field at the triple points near the electrodes, where the electron cascade is most likely to initiate~\cite{wetzer1995hv, yamamoto1997numerical}. For example, experiments in liquid argon have found that surface discharges can span more than \SI{60}{\centi\meter} of ribbed insulator length at relatively low average fields (\SI{<20}{\kilo\volt\per\centi\meter}) when initiated at a triple point with a maximum field of $\sim$\SI{200}{\kilo\volt\per\centi\meter}~\cite{blatter2014experimental}.

\subsection{Liquid thermodynamics and purity}
Several studies of high voltage in noble liquids have hypothesized that breakdowns are initiated by localized heating of the liquid causing vapor formation~\cite{phan2021study, watson2023study}. The lower permittivity of the gas tends to enhance the electric field in the vapor bubble~\cite{Babaeva2009structure} and the dielectric strength of noble gases is significantly lower than that of noble liquids. Bubbles in high field regions can therefore initiate breakdowns. The visual appearance of bubbles has been correlated with the onset of breakdowns in liquid helium~\cite{okubo1996high}, argon~\cite{bay2014evidence, blatter2014experimental}, and xenon~\cite{watson2023study}. Near atmospheric pressure, noble elements have a relatively narrow temperature range over which they are liquid. One should ensure that there are no localized heat sources (such as resistors, electronics, etc.) in high field regions that can cause boiling~\cite{breur2021measurement}. Operating at elevated pressure and reduced temperature, away from the liquid-gas saturation curve, can suppress both bubble nucleation and growth, with experiments measuring a significant increase in breakdown voltage with increasing pressure in liquid helium~\cite{phan2021study}.  

The presence of electronegative impurities, such as oxygen, in noble liquids increases the probability of electron attachment, thereby decreasing the probability of ionizing collisions that initiate breakdown~\cite{swan1961influence}. The electric field at which breakdown occurs in liquid argon has been observed to increase with increasing impurity levels~\cite{blatter2014experimental, acciarri2014liquid}, but this effect is normally subdominant to other parameters such as electrode size and geometry. Current and next-generation experiments typically require the target noble liquid to have electronegative impurities below \SI{0.1}{\ppb} (oxygen-equivalent) for efficient drifting of ionization electrons across meter-scale active volumes. Test setups for HV components should therefore also aim for similar impurity levels, as higher impurity concentrations might lead to deceptively high breakdown fields. 

\section{HV experience from EXO-200}
\begin{figure}
    \centering
    \includegraphics[width=\linewidth]{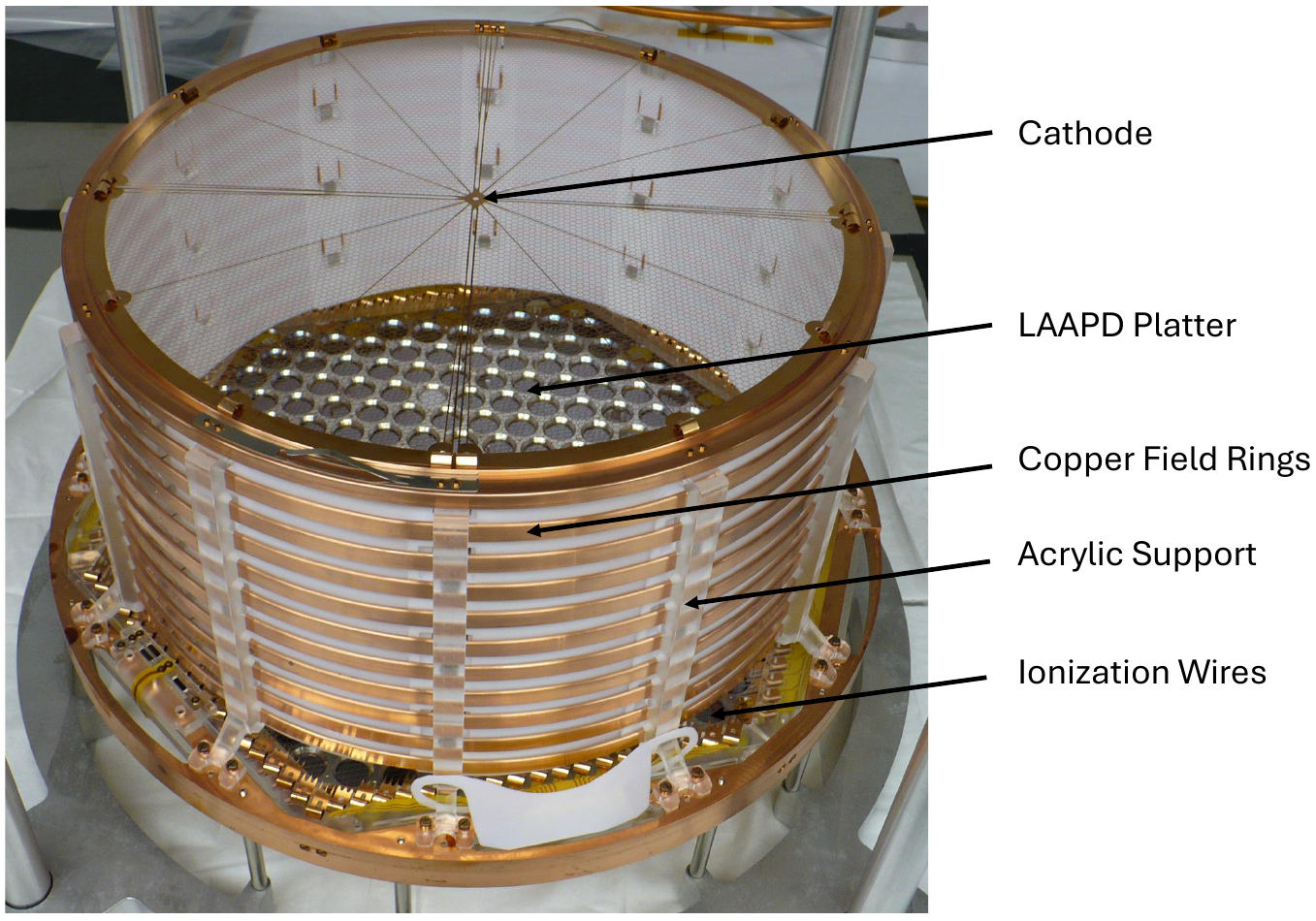}
    \caption{One half of the EXO-200 double TPC. The etched phosphor bronze cathode grid at the top in the photo also serves as the cathode for the other half of the TPC which is symmetrically placed above. Note the PTFE panels on the interior wall of the field cage, and the \ce{Al}/\ce{MgF2} coating of the copper LAAPD platter, both designed to improved reflectivity for VUV light.}
    \label{fig:exo200}
\end{figure}
The design of the nEXO TPC and high voltage system is based on the EXO-200 HV R\&D program (2011-2018) and the experience of operating the EXO-200 detector~\cite{exo2019search} near Carlsbad New Mexico at the Waste Isolation Pilot Plant (WIPP). The EXO-200 TPC~\cite{exo2012exo200} was a cylindrical ``double TPC'' with a cathode grid held at negative high voltage in the middle and two independent charge and light readout detection planes near ground potential at each end (see Fig.~\ref{fig:exo200}). The drift region of each individual TPC was \SI{19.8}{\cm} in length and \SI{18.3}{\cm} in radius. Ionization charge was detected at the anodes through two tensioned planes of etched phosphor bronze wires, read out to measure both induction and collection signals. Scintillation light was detected by an array of large area avalanche photodiodes (LAAPDs) behind each charge collection plane. 

The EXO-200 power supply output was filtered by two sets of RC filters. The second filter included a capacitively-coupled connection to an oscilloscope allowing for the detection of small high frequency voltage fluctuations (referred to as ``glitches''). The voltage bias was delivered to the cathode by a HV coaxial cable contained within copper tubing that was an extension of the copper xenon vessel, containing both liquid xenon (LXe) and, at the warm end \SI{\sim 1.5}{\meter} away, gaseous xenon, where there was an o-ring seal to the cable.  The HV cable plug engaged with a sprung receptacle connected to the cathode plane. The voltage was stepped down by ten circumferential field shaping rings connected to voltage-divider resistors, with the rings supported and separated by longitudinally positioned acrylic ``combs''. There was a radial clearance of \SI{11}{\milli\meter} between the rings (at voltage) and the copper vessel (at ground), with the acrylic combs extending \SI{8}{\milli\meter} radially out from the copper rings.

The EXO-200 TPC was designed to work up to a cathode voltage of \SI{75}{\kilo\volt} corresponding to a drift field of \SI{3700}{\volt\per\cm}~\cite{exo2012exo200}. During the initial ramp up of the EXO-200 HV system, the frequency of small glitches (\SIrange{2}{20}{\milli\volt} in amplitude) on the HV delivery line started to increase beyond a cathode voltage of \SI{-9}{\kilo\volt}. To avoid the risk of damage from a discharge before a full dataset could be collected, the detector was initially operated at a cathode voltage of \SI{-8}{\kilo\volt}, corresponding to a drift field of \SI{380}{\volt\per\cm}.

A mock-TPC was built to better understand the factors limiting stable HV operation. It differed from EXO-200 only in the diameter of the field rings (\SI{19}{cm} OD, half of EXO-200) and the use of stainless steel conductors. A series of tests were performed in a LXe cryostat equipped with cryogenic cameras to study the HV stability of the mock-TPC as a function of the applied voltage~\cite{Delaquis2015}. The system was able to ramp up to \SI{-38}{\kilo\volt} before discharges were observed. Images provided by the cryo-camera revealed the location of the spark, thanks in part due to phosphorescence in the acrylic supports for the field shaping rings. The discharges occurred between the field shaping ring at the highest voltage and a grounded vessel wall \SI{11}{\milli\meter} away (same distance as in EXO-200) and likely originated at a triple-point where the conductor, acrylic, and LXe were in contact. FEA simulations using COMSOL\textsuperscript{\textregistered}~\cite{comsol2025} confirmed the presence of a locally enhanced field at the triple-point location, with the maximum electric field in the liquid xenon reaching roughly \SI{90}{\kilo\volt\per\cm} at a cathode voltage of \SI{40}{\kilo\volt} (see Sec. 6.3 of Ref.~\cite{Delaquis2015} for more details). Similar findings were obtained during HV testing of the actual EXO-200 detector after the first two years of data-taking. On ramping the voltage beyond \SI{-8}{\kilo\volt}, small discharges were detected that appeared to be occurring at locations close to the acrylic supports, based on signals observed by LAAPDs at large radius. Despite an increased rate of glitching during ramping, the system ran stably at \SI{-12}{\kilo\volt} throughout the test. 

As a result of these investigations, the cathode voltage was raised from \SI{-8}{\kilo\volt} to \SI{-12}{\kilo\volt} during operation with the upgraded EXO-200 detector (Phase II)~\cite{exo2018search}, increasing the electric drift field to \SI{567}{\volt\per\cm} and contributing to improved energy resolution. The detector operated stably at this voltage for the remaining 2.5 years of continuous operation with small glitches occurring only a few times a month and no glitches with an amplitude greater than \SI{2}{V} detected. Prior to decommissioning EXO-200 in 2018, a final HV test was performed where the voltage was raised to \SI{-25}{\kilo\volt}. Glitches were observed during the tests, but no full discharge occurred, consistent with the mock-TPC tests. 

Based on the EXO-200 experience it is clear that detailed electrostatic simulations are needed at the early design stages and that special attention should be paid to recessing and shielding triple point junctions. Additionally, the mock-TPC tests emphasize the need for testing HV components in as close to final configuration as possible.

\section{nEXO HV delivery}
A common component of TPC HV design is the connection between the HV power supply and the electrode (typically the cathode, biased at negative voltage with respect to ground) inside the TPC\footnote{There are some prototype designs that generate the high voltage internally~\cite{Ereditato2013design, romero2020avolar, akiyama2025insitu}.}. 
The connection at the electrode is typically the region that contains the highest electric fields and stressed area, and is therefore the region where the design considerations discussed in the previous section are most relevant. 
In this section we present a design concept for a HV delivery system in the context of the nEXO detector that takes into account the previously mentioned design guidelines while meeting the strict radiopurity and space constraints. 

\subsection{nEXO overview}
nEXO is a concept for a next-generation liquid xenon experiment searching for neutrino-less double beta decay of \isotope{Xe}{136}~\cite{nexo2018sensitivity}. %The nEXO detector design takes advantage of the properties of liquid xenon (LXe), such as  i) high detection efficiency, ii) \tbd{gamma self-shielding} (due to high-Z and high density), iii) ability to maintain (initially and over the course of the experiment) extremely high chemical and radioactive purity. 
At the center of the experiment is a single-phase TPC enclosed in an ultra-low background copper~\cite{HOPPE2014116} vessel containing \SI{\sim4.8}{\tonne} of LXe isotopically enriched to \SI{90}{\%} \isotope{Xe}{136}. The TPC drift region is a monolithic \SI{1.13}{\meter} OD and \SI{1.18}{\meter}-long cylindrical volume, with \SI{3.65}{\tonne} of LXe contained within an open field cage with the anode and the cathode at opposite ends. Ionization electrons and scintillation photons from energy deposits in the drift region are recorded respectively with a segmented charge-sensitive anode~\cite{jewell2018characterization} and a large ($\sim$\SI{4}{\square\meter}) VUV-sensitive silicon photomultiplier (SiPM) array~\cite{gallina2022performance} covering most of the surface of the cylindrical vessel barrel. The copper vessel is cooled to liquid xenon temperatures by immersion inside a low-background double-walled nickel~\cite{roosendaal2025ultra} cryostat filled with a radiopure liquid hydrofluoroether (HFE) refrigerant (1-methoxyheptafluoropropane, \ce{C3F7OCH3}, trade name Novec 7000~\cite{novec7000}). The HFE vessel is contained in a large instrumented tank (\SI{12.3}{\meter} in diameter and \SI{12.8}{\meter} tall) filled with \SI{1.5}{\kilo\tonne} of ultra-pure deionized water that serves as a radiation shield and water Cherenkov muon detector.

\subsection {nEXO HV system overview}
\begin{figure}
    \centering
    \includegraphics[width=1\linewidth]{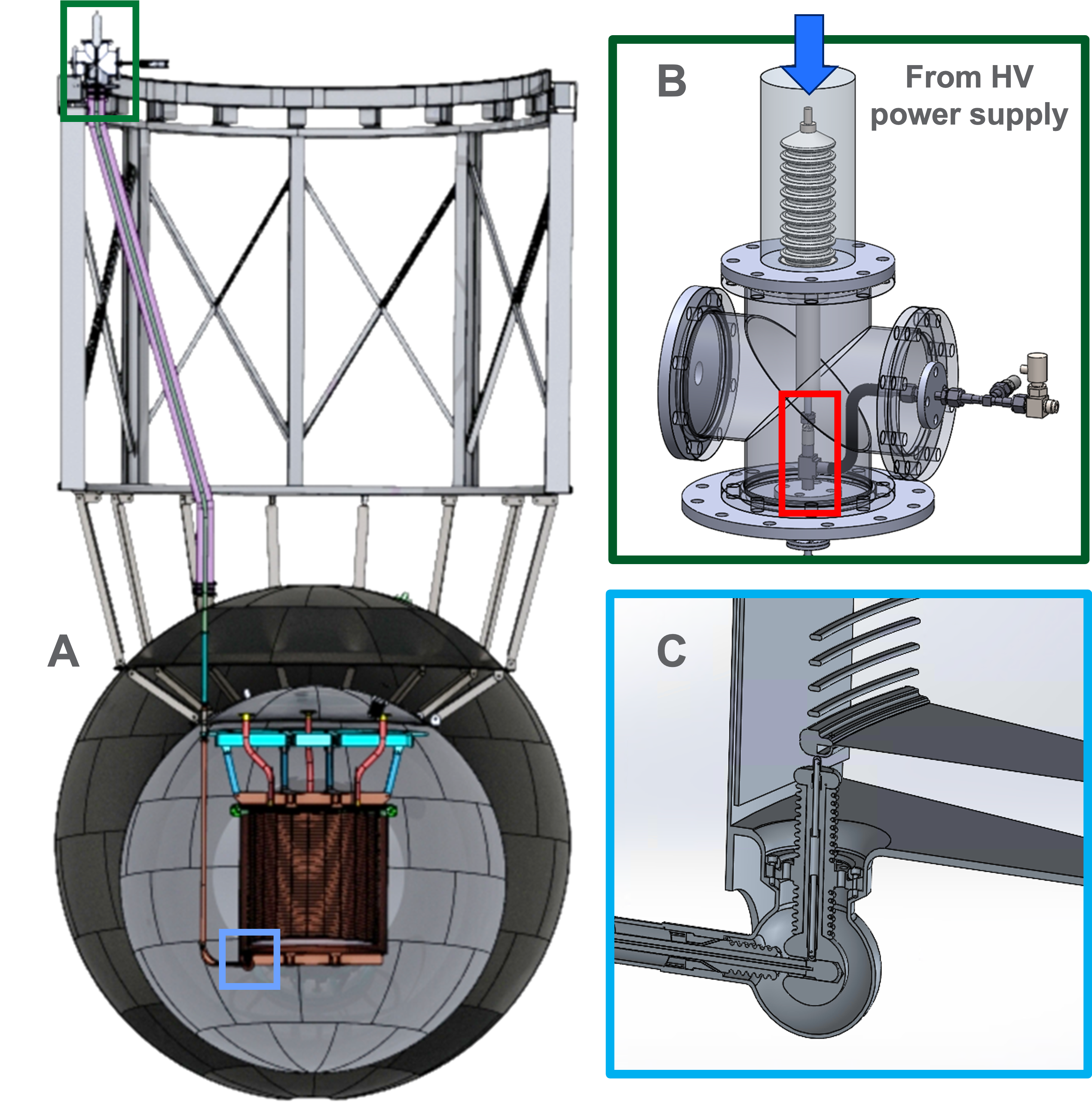}
    \caption{A: Overview of nEXO HV delivery system. It consists of a HV power supply and filter (not shown) that connects to a pressure-rated HV-feedthrough (inset B, with more details in Fig.~\ref{fig:warm_seal}) which contains a room-temperature xenon-air seal around a continuous HV cable. The cable runs through a conduit that passes through the water tank and HFE cryostat and connects at the bottom of the TPC vessel (inset C) with a HV connection which delivers voltage to the cathode.}
    \label{fig:hv_system_overview}
\end{figure}

The design of the nEXO HV and Field Cage system, shown in Fig.~\ref{fig:hv_system_overview}, is similar to that of the EXO-200 experiment~\cite{PhysRevLett.107.212501, PhysRevLett.109.032505} and other liquid xenon-based TPCs and is based on an HV R\&D program that ran between 2011 and 2018 within the EXO-200 collaboration. The electric field is generated by applying a negative voltage to the cathode located near the bottom of the TPC and connecting the anode at the top of the TPC to ground. Voltage is supplied by a commercial power supply that sits on the external deck. The output of the power supply is filtered to reduce noise and monitored as an early-warning system for HV instability. A continuous commercial HV cable then carries the bias from the external deck through the water tank and cryostat vessels to the bottom of the TPC where it connects to the cathode. The cable is contained in a narrow sealed conduit to protect the cable from the water and HFE. The placement of the photosensors along the barrel of the TPC vessel allows for the use of a solid plane cathode, unlike wire grids typically used in experiments with photosensors below the cathode. The direction of the electric field is kept uniform by a field cage, made of coaxial copper field shaping rings (FSRs) connected by resistors that grade the potential between the cathode and anode. The FSRs and cathode are coated with a VUV reflector (aluminum protected by magnesium fluoride) to enhance the collection of xenon scintillation light. 

The target electric field magnitude for the nEXO TPC is driven by maximizing the sensitivity of the experiment to neutrinoless double beta decay. Considering a range of factors that vary with electric field such as electron-ion recombination, electron drift velocity, electron attachment to impurities, diffusion of drifting ionization electrons, distortion of field uniformity by surface charge build-up, and standoff distance between HV components and the grounded vessel, it was determined that for the given nEXO TPC volume the optimal operating electric field is \SI{\sim400}{\volt\per\centi\meter}. This corresponds to an applied voltage on the cathode of approximately \SI{-50}{\kilo\volt}. Moreover, it was found that the nEXO sensitivity was relatively independent of small changes in the operating field, decreasing by only \SI{\sim 5}{\%} when operating at a field of \SI{200}{\volt\per\centi\metre}, though it decreases more sharply at lower operating fields. Given these findings, the design goal for the nEXO HV system is to operate stably with a cathode voltage of \SI{-50}{\kilo\volt}.

The radioactive background requirements needed to reach a neutrinoless double beta decay half-life sensitivity of more than \num{1E+28} years correspond to a target effective background index of \SI{7E-5}{\counts\per\FWHM\per\kilo\gram\per\year}~\cite{adhikari2022nexo}. This imposes very stringent requirements on the radiopurity of all detector materials, especially those close to the active TPC volume. The entire HV delivery system was allocated \SI{\sim 4}{\%} of the total background budget, which restricted the design to use only extremely radiopure materials and minimize the mass of all components as much as possible. For all conductive components we chose either copper or alloys of copper due to the availability of radiopure copper, despite their non-optimal high voltage properties. For insulating and semi-conducting material, we chose polyethylene-based materials, minimizing the mass as much as possible in favor of using the liquid xenon itself as the insulating medium. Tab.~\ref{tab:radiopurity_summary} shows radiopurity measurements of the materials chosen for the nEXO HV delivery system to meet the allocated radioactivity budget.

\begin{sidewaystable}
    \begin{center}
        \def\arraystretch{1.2}  
        \begin{tabular}{lllccccr}
            Material & & HV component & \isotope{U}{238} (\unit{\ppt}) & \isotope{Th}{232} (\unit{\ppt}) & \isotope{K}{nat} (\unit{\ppb}) & \isotope{Rn}{222} (\unit{\micro\becquerel\per\meter}) & Reference \\
            \hline
            \multirow{2}{*}{Copper} & Electroformed & angled connection & \num{<9.4e-3} & \num{6(1)e-3} & & & R-168.1.1~\cite{nEXORBC}\\
                            & Commercial & cable conduit & \num{0.25(0.01)} & \num{0.13(0.06)} & \num{<0.7} & & R-002.11.1, R-002.12.1~\cite{nEXORBC} \\
            \hline
            Phosphor Bronze         & & cable pill, screws, springs & \num{<0.35} & \num{<2.3} & \num{0.22(0.05)} & & R-208.1.1~\cite{nEXORBC} \\
            \hline
            \multirow{4}{*}{DS Cable (2014)} & Layers A,B,C & HV cable & \num{100(40)} & \num{64(7)} & \num{350(40)} & \num{<240} & R-021.4.1, R-021.9.1~\cite{nEXORBC}\\
                              & Layer A & cable core & \num{440(200)} & \num{370(40)} & \num{1990(220)} & & R-021.5.1~\cite{nEXORBC} \\
                              & Layer B/LDPE & cable insulator & \num{<110} & \num{18.6(2.0)} & \num{132(14)} & & R-021.6.1~\cite{nEXORBC} \\
                              & Layer C & cable ground & \num{760(330)} & \num{900(100)} & \num{2690(300)} & & R-021.7.1~\cite{nEXORBC} \\
            \hline
            DS Cable (2023) & Layers A,B,C & HV cable & \num{<580} & \num{900(500)} & \num{3000(600)} & \num{<8.1} & R-206.1.1, R-206.2.1~\cite{nEXORBC, nEXOESC} \\
            \hline
            \multirow{2}{*}{Polyethylene} & UHMW PE & \multirow{2}{3.5cm}{stress cone} & \num{65 \pm 24} & \num{<340} & \num{<126} & & 108~\cite{xenon2017material} \\
                        % & LDPE (same as Layer B) & & \num{<106} & \num{18.6(0.7)} & \num{132(4)} & & R-021.6.1 \\
                        & Tivar ESD & & \num{300(130)} & \num{730(260)} & \num{820(270)} & & R-201.1.1~\cite{nEXORBC}\\
            \hline
        \end{tabular}
        \caption{Summary of the material radiopurity for the components in the nEXO HV delivery system. The contamination levels for \isotope{U}{238}, \isotope{Th}{232} and \isotope{K}{nat} are normalized by mass while the \isotope{Rn}{222} column lists the steady state radon emanation activity per unit length of cable.}
        \label{tab:radiopurity_summary}
    \end{center}
\end{sidewaystable}

\subsection{HV power supply and filters}
In the nEXO TPC design, which does not contain a shielding (Frisch) grid, the cathode is capacitively coupled to the charge readout channels at the anode. Any fluctuations in the cathode voltage will induce signals in the charge readout, so the requirements for voltage stability are very stringent, particularly in the sensitive frequency range of the charge preamplifiers. Most power supplies generate high voltage from an RF oscillator at \SIrange{30}{100}{\kilo\hertz} and then step it up using a transformer. This frequency lies in the center of the passband of the nEXO front end electronics. Calculations showed that the coupling capacitance of the cathode to an individual charge channel is about \SI{5}{\femto\farad}, so \SI{1}{\volt} of noise corresponds to about \SI{30000}{electrons} of charge, which would pass unfiltered through the signal chain. To meet the requirements of less than \SI{30}{electrons} of charge-equivalent noise from the cathode HV supply, the nEXO TPC needs a high voltage supply capable of \SI{-50}{\kilo\volt} output with a noise level of well under a millivolt.

\begin{figure}
    \centering
    \includegraphics[width=1\linewidth]{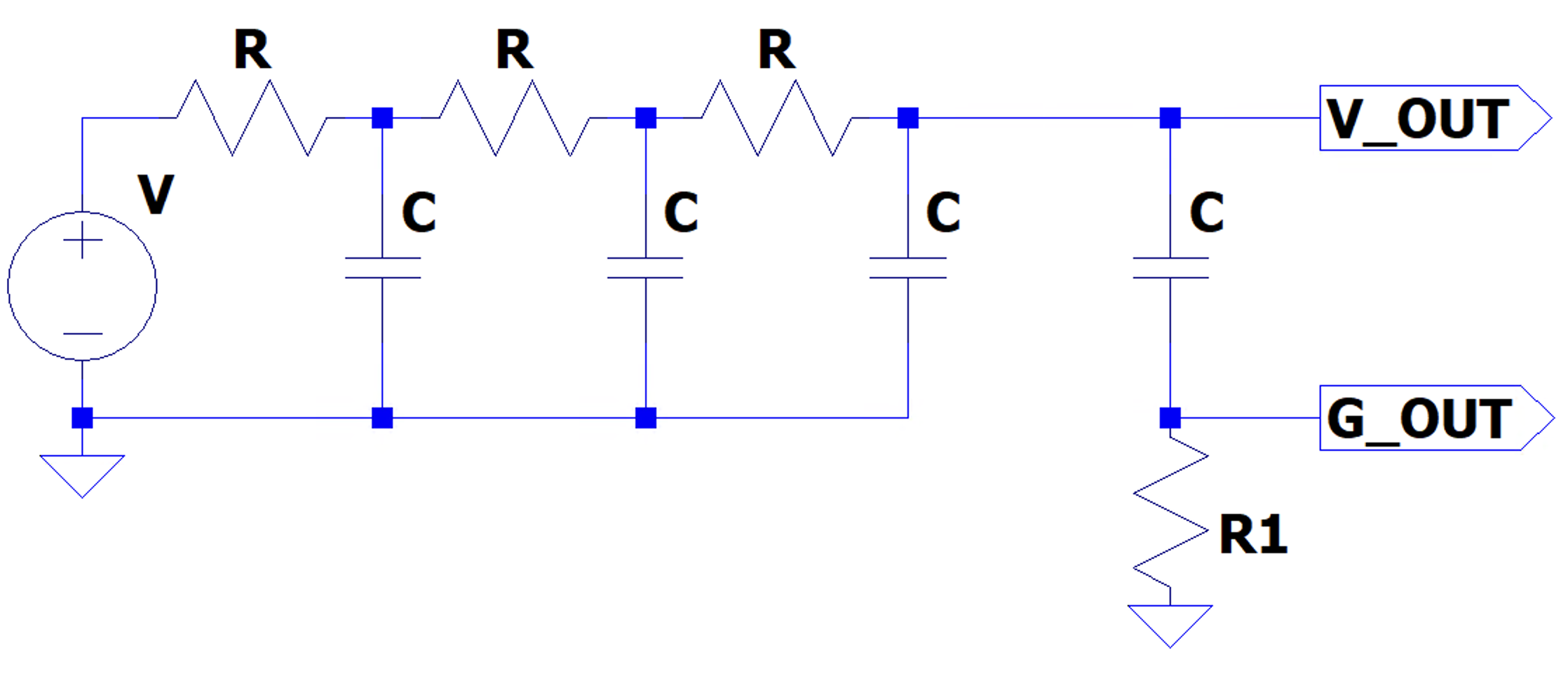}
    \caption{Circuit diagram of the HV power supply, noise filter, and glitch detector system. $V_{out}$ connects to the TPC cathode through a HV cable and $G_{out}$ connects to the oscilloscope for the glitch detector. $R = \SI{10}{\mega\ohm}$, $C = \SI{1}{\nano\farad}$, and $R_1 = \SI{1}{\mega\ohm}$. See text for details.}
    \label{fig:hv_filter}
\end{figure}

The supply chosen for this work, a Spellman SL70N30~\cite{SpellmanPS}, produces up to \SI{-70}{\kilo\volt} but has a triangle-shaped output noise with an amplitude of about \SI{10}{\volt} at a current draw of a few microamps. Therefore the high voltage must be filtered by at least a factor of \num{10000} to reach the target charge-equivalent noise. This was accomplished using a 3-stage RC filter (see Fig.~\ref{fig:hv_filter}) consisting of \SI{10}{\mega\ohm} resistors and \SI{1}{\nano\farad}, \SI{100}{\kilo\volt} capacitors~\cite{HVCAP}. Each stage has a time constant of \SI{10}{\milli\second} and a nominal noise attenuation factor of several hundred. It was found that the limiting factor of the filter performance was poor shielding from one filter stage to the next, so the filters were placed in three separate enclosures. The enclosures were made from NW-200 flanged cylinders filled with silicone oil~\cite{SuperlubeOil} to prevent breakdown. The resulting noise level achieved was less than \SI{50}{\micro\volt} RMS at \SI{35}{\kilo\hertz}. We note that in the nEXO detector configuration the additional resistance and capacitance of \SI{\sim 10}{\meter} of HV cable between the power supply and cathode will add additional filtering (see Sec.~\ref{sec:cable}).

An additional feature of the Spellman SL70N30 is that it has a floating ground so that the current draw of the resistor chain in the TPC can be monitored. This can detect open or shorted field electrodes during installation or operation. It was tested with a \SI{10}{\kilo\ohm} monitoring resistor and showed \SI{\sim10}{\nano\ampere} of resolution, much better than required for the resistor chain, which is expected to draw several microamps.

Based on the HV testing done for EXO-200 described above, the filter system also includes a ``glitch'' detector to monitor and detect voltage fluctuations which were observed to precede HV breakdown. These glitches typically start at only a few millivolts of amplitude but grow exponentially near breakdown, with the rate increasing to several \unit{\hertz} and the amplitude exceeding several hundred millivolts. The glitch detector can thus serve as an early warning system for high voltage instability and allow for the ramping down of the high voltage before a full discharge occurs \cite{elbeltagi2024thesis}. This is especially important in the nEXO design, where a discharge could damage the light and charge readout electronics inside the TPC. The glitch detector consists of a \SI{1}{\nano\farad} high voltage capacitor with one end connected to the high voltage line and the other end to a 16-bit USB oscilloscope with microvolt sensitivity~\cite{PICOScope}. The oscilloscope for glitch detection has built-in triggering and a recording function so that voltage traces of glitches can be stored and analyzed offline. The trigger threshold for glitches was usually set at \SIrange{2}{5}{\milli\volt}.

A test system consisting of the HV supply, filters, and glitch detector was operated at \SI{-70}{\kilo\volt} (higher than the nEXO design voltage of \SI{-50}{\kilo\volt}) for several thousand hours. The recorded glitch rate was zero to a few per week, with no breakdowns observed.

\subsection {HV cable}
\label{sec:cable}
\begin{figure}
    \centering
    \includegraphics[width=1\linewidth]{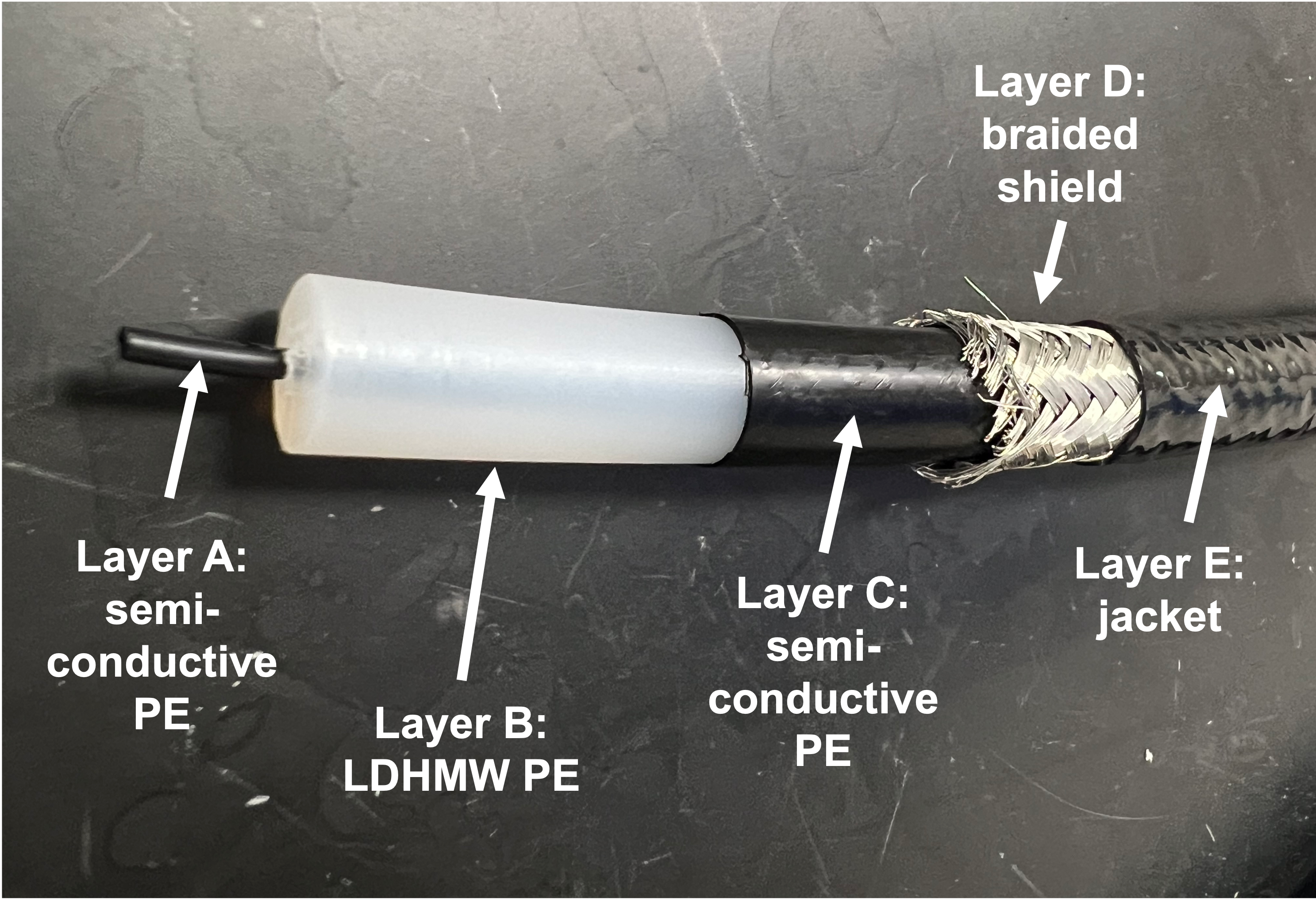}
    \caption{High voltage cable from Dielectric Sciences, Inc. (cable DWG NO. 2353). From the center moving outwards, the cable consists of: Semi-conductive PE conductive core (Layer A); LDHMW PE insulator (Layer B); Semi-conductive PE ground sheath (Layer C); Braided shield (Layer D); Outer jacket (Layer E).}
    \label{fig:dscable}
\end{figure}

The HV cable is a commercially available all-polyethylene (PE) coaxial cable from Dielectric Sciences, Inc (DWG NO. 2353)~\cite{DieletricSciencesInc}. It features a semi-conductive polyethylene conductor and ground sheath with a standard low density high molecular weight polyethylene (LDHMW PE) insulator (see Fig.~\ref{fig:dscable}).
The all-polyethylene construction eliminates concerns about differential thermal contraction between the central conductor (typically metal) and insulator and also reduces the heat conduction through the conductor.
PE in its different forms (LDPE, LDHMW PE, XLPE) is a good material for HV applications in a cryogenic environment because it has excellent dielectric properties, provides effective electrical insulation, and maintains better flexibility at low temperature than other plastics such as PVC~\cite{ware1967polyethylene, zhou2017polymeric}.

The DS-2353 cable is rated for operation at \SI{150}{\kilo\volt} and was extensively tested in liquid xenon as part of the EXO-200 HV R\&D program, leading to this cable being deployed by the LZ collaboration~\cite{akerib2019lz} and the DUNE ND experiment~\cite{abuddune2021}, with a similar cable intended for use in the DarkSide-20k~\cite{UCD2021} experiment. We note that samples purchased over the period from 2014-2023 exhibited some variability depending on the production batch (physical dimensions varied by a few percent, while electrical properties, such as the resistivity of the conductive core, varied by a factor of $>2$). The cable used in the tests described here was purchased in 2023 and has the following characteristics (see Fig.~\ref{fig:dscable} for layer labels): the core conductor (layer A), has a diameter of \SI{2.0}{\milli\meter} (\SI{0.08}{in}), while the insulator (layer B) extends to a diameter of \SI{11.2}{\milli\meter} (\SI{0.44}{in}) with a ground sheath (layer C) of \SI{0.25}{\milli\meter} (\SI{0.01}{in}) wall thickness.
The cable comes with a braided shield (34AWG tinned copper, \SI{95}{\%} coverage --- layer D) and a polyester-based polyurethane jacket (\SI{12.7}{\milli\meter} (\SI{0.5}{in}) diameter --- layer E), both of which are removed for our application. The conductive core has a resistance of \SI{20}{\kilo\ohm\per\meter}, and the cable has a capacitance of \SI{73}{\pico\farad\per\metre}. 

In order to relieve internal stress present in the cable from the cable fabrication process (co-extrusion of the various layers) and straighten it out following its shipment and storage in a \SI{12}{in} (\SI{305}{\milli\meter}) diameter coil, an annealing procedure was developed in collaboration with colleagues at University of California, Davis~\cite{UCD2023}. For some batches of cables, annealing also prevented the risk of fracturing when rapidly cooled to liquid nitrogen temperatures~\cite{UCD2023}. The cable annealing was performed using custom designed ovens to thermally cycle individual sections of the cable at a time. Our annealing procedure, based on commercial annealing procedures for LDPE~\cite{BoedekerPlasticAnnealing}, starts with a section of the cable held straight at room temperature. It is then heated to \SI{80}{\degreeCelsius} with a ramp rate of \SI{+0.5}{\degreeCelsius\per\min}. The cable soaks at \SI{80}{\degreeCelsius} for $1$\,h, and is subsequently cooled down to room temperature with a ramp rate of \SI{-0.1}{\degreeCelsius\per\min}. While the annealed cable is visually identical to the pre-annealed state (except being straighter) it has a noticeably lower bending stiffness, its physical dimensions changed slightly (\SI{<1}{\%}), and the electrical resistance of the core increased by \SI{\sim 25}{\%}. The change in resistivity is likely related to rearrangement of the filler particles forming the conductive network in the polyethylene matrix~\cite{liang2013, liang2017}.

To verify that it meets the nEXO radiopurity requirements, the cable was screened for radioactive contamination. Using samples from an early version of the cable, purchased in 2014, measurements of the radiopurity of the whole cable, as well as each of the individual layers (A, B, C), were made using neutron activation analysis (NAA). From the most recent batch purchased in 2023, \SI{10}{\meter} (\SI{33}{ft}) of cable (\SI{1049}{\gram}), was cut into \SI{6}{in} (\SI{152}{\milli\meter}) sections for HPGe gamma counting~\cite{nEXORBC}. A separate \SI{10}{\meter} section of cable was measured for radon outgassing~\cite{nEXOESC}. The assay results are summarized in Tab.~\ref{tab:radiopurity_summary}.

With an all polymer construction, the cable can act as a source of outgassing, potentially compromising the LXe chemical purity.  Measurements of the solubility and diffusion coefficient of oxygen in polyethylene~\cite{LZTDR} imply that the oxygen content in the cable can be reduced to a negligible level by vacuum pumping the cable for 3-4 days and then storing in an inert atmosphere. Even without prior outgassing, assuming that the diffusion follows the Arrhenius equation, measurements of the activation energy of oxygen diffusion in polyethylene ($E_a = \SI{10}{\kilo\cal\per\mole}$~\cite{tochin1975low}) predict a maximum outgassing rate of \SI{<2E-9}{\gram\per\meter\per\day} for the sections of the cable at LXe temperature near the TPC, well below the purity requirements for nEXO. Finally, simplified laminar-regime fluid simulations indicate that even a small flow rate (\SI{\sim 1}{SLPM}) of xenon in the cable conduit away from the TPC is sufficient to make the outgassed contamination entering the active TPC region negligible compared to nEXO requirements. If needed, this flow of xenon can be facilitated by connecting the conduit, below the O-ring seal, to the gas recirculation and purification system.

\subsection{Room temperature gas seal}
The seal of the xenon volume in the conduit around the HV cable is made at room temperature outside of the water tank. Designs with the location of a cryogenic xenon volume seal close to the TPC vessel were initially considered, but the placement of the seal far from the active volume allows the use of standard pressure/vacuum rated fittings and high voltage feedthroughs without concern for radiopurity. Additionally, the seal is at room temperature, which eliminates issues related to mismatched thermal expansion and cryogenic compatibility of sealing materials. An accessible seal also allows one to monitor for leakage of xenon through the seal. 

\begin{figure}
    \centering
    \includegraphics[width=1\linewidth]{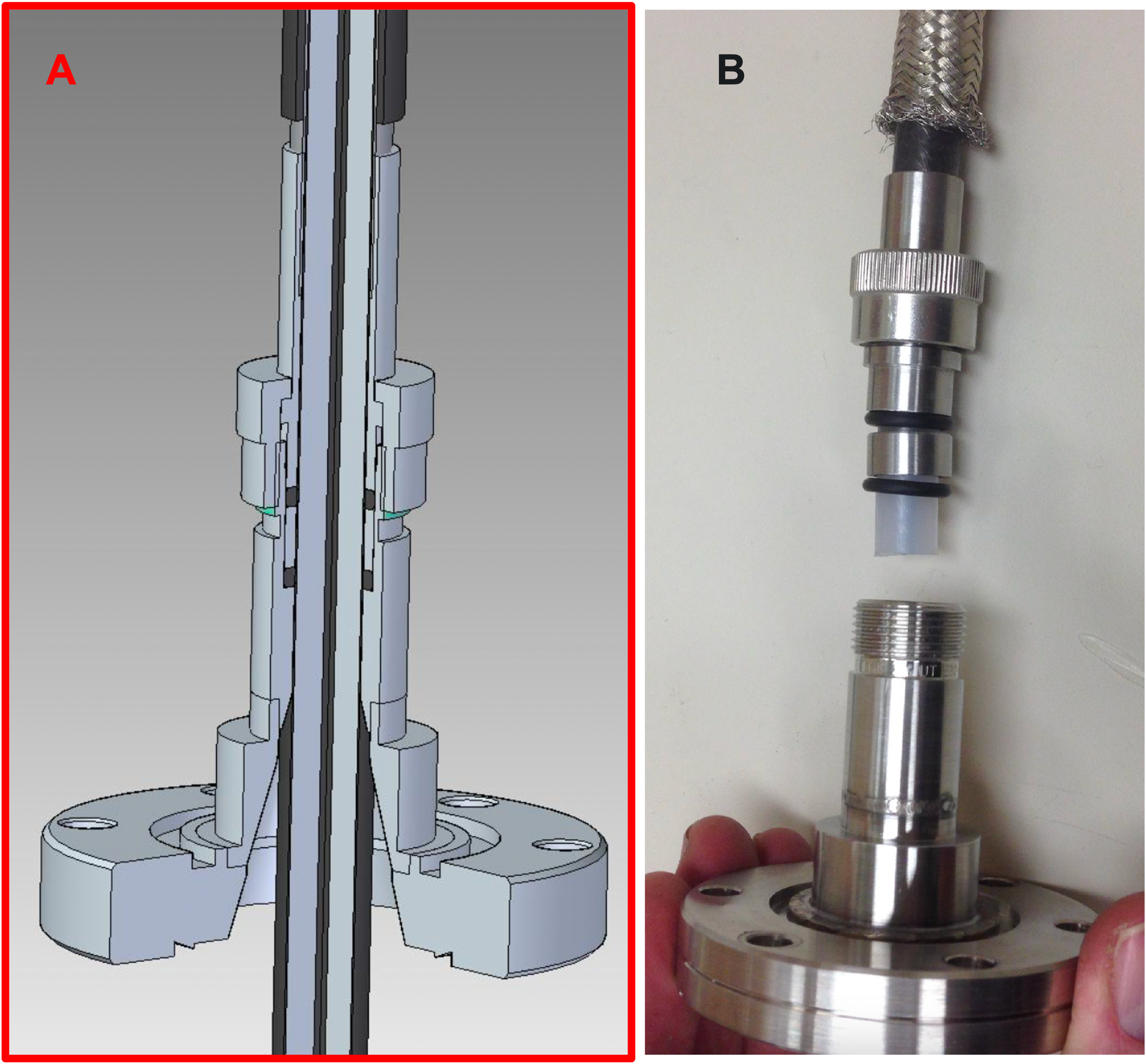}
    \caption{A: Conceptual model of the room temperature gas seal around the HV cable on a customized \SI{2.75}{in} (\SI{69.8}{\milli\meter}) CF flange. B: A prototype built and tested in the lab showing the modified Swagelok\textregistered~Ultra-Torr fitting, and the double O-ring.}
    \label{fig:warm_seal}
\end{figure}

The cable feedthrough uses a modified Swagelok\textregistered~Ultra-Torr fitting~\cite{UltraTorrFittings} with a butyl (chosen for its low permeability) O-ring (size 906) to seal around the conductive ground layer (Layer C) of the cable. It features an additional O-ring that serves as both a strain relief for the cable that extends through the conduit below and as a redundant seal (see Fig.~\ref{fig:warm_seal}A), acting as an intermediate space to trap any leaks caused by movement of the joint. Fig.~\ref{fig:warm_seal}B shows a version of this warm feedthrough made by modifying a commercially available Ultra-Torr to Conflat (CF) feedthrough to accept the additional O-ring. The prototype was successfully leak tested to have a He leak rate \SI{<2E-8}{\milli\bar\liter\per\second}. Similar versions of this design have been installed in the LZ experiment~\cite{LZTDR} and the DUNE Near Detector~\cite{abuddune2021}.
It should be noted that if the experiment grounding scheme requires the detector high voltage components to be electrically isolated, the pass-through can be coupled to a ceramic break or the internal surfaces can be coated with an insulator to isolate the cable ground layer from the fitting.

In order to meet standard pressure regulatory requirements, the cable pass-through seal is housed in an external pressure-rated vessel (Swagelok\textregistered~Ultra-Torr fittings are only rated for vacuum). The vessel can be maintained at vacuum to allow for monitoring of xenon leakage through the cable feed-through seal using a residual gas analyzer (see for example Ref.~\cite{LZTDR}).

The conduit around the cable connects the deck level gas seal to the high voltage connection at the bottom of the TPC as a continuous fluid volume (see Fig.~\ref{fig:hv_system_overview}). It is specifically designed to facilitate the insertion or removal of the cable from the deck level after assembly of the TPC is complete. It is also designed to accommodate thermal contraction of the cable as it is cooled down from room to liquid xenon temperature through the use of flexible bellows and an expansion volume. Outside the cryostat the conduit sits in a vacuum-insulated line, and the transition between xenon liquid and gas in the conduit is set by the height of the main xenon condensation system.

\subsection{Cable ground termination}
In order to connect a coaxial HV cable to an electrode or other component the outer conducting ground layer must be terminated. Without the application of any electric field grading, there is typically a very high electric field enhancement at the edge of the terminated ground layer~\cite{ye2018review} that can serve as the initiation point for surface flashovers and high voltage breakdowns~\cite{blatter2014experimental}. The most common technique to mitigate the electric field at a ground termination is to use a stress cone that fits over the cable at the location of the ground termination. There are many types of stress cones and field grading techniques~\cite{ye2018review}, though here we focus on the geometric electrode grading technique which is commonly used for polymer cables. 

\begin{figure}
    \centering  \includegraphics[width=1\linewidth]{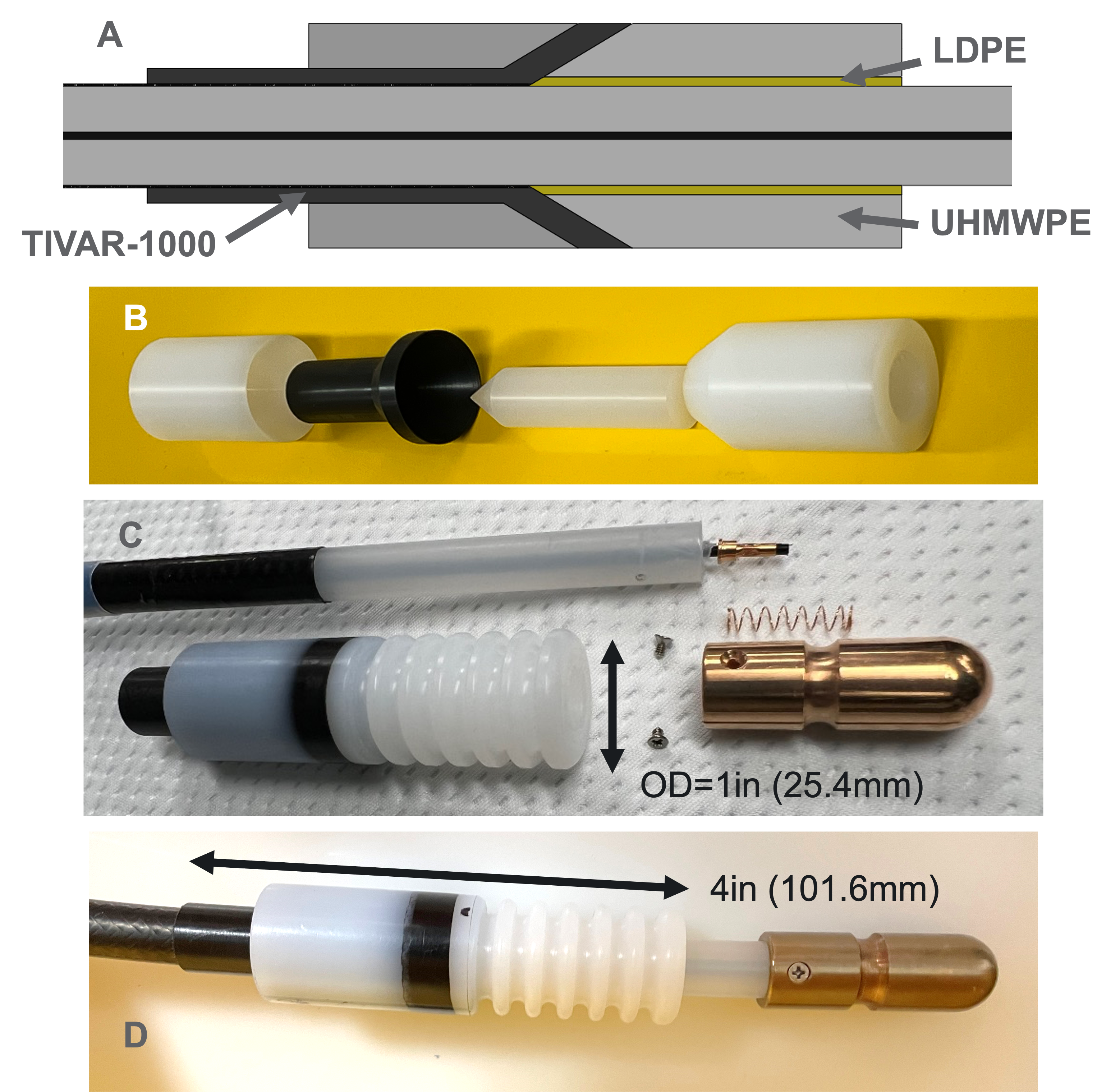}
    \caption{Stress cone design and prototypes (A) Cross-section of stress cone around the HV cable showing the different layers. (B) Machined stress cone components before thermal bonding. (C) Components of the HV cable conductor termination. (D) Complete HV cable termination with stress cone bonded onto the cable and conductive pill attached.}
    \label{fig:stress_cone}
\end{figure}

Fig.~\ref{fig:stress_cone} shows the proposed stress cone design based on the considerations discussed in Sec.~\ref{sec:insulators}. The purpose of this stress cone is two-fold: (a) to reduce the electric field at the ground termination by gradually separating ground away from the cable and (b) to displace the liquid xenon from the high electric field in this region using insulating polymers with higher dielectric strength. 

The proposed stress cone design is composed of several polyethylene layers (see Fig.~\ref{fig:stress_cone}) which are thermally bonded together and to the cable. Bonding the stress cone to the cable ensures that xenon cannot enter the high field region between the stress cone and the cable. Ultra-high molecular weight polyethylene (UHMWPE) is used for the majority of the insulating component while semi-conductive UHMWPE (Tivar-1000 ESD) is used to electrically connect to the cable ground (layer C) and serve as the deflector to gradually separate the ground potential and reduce the electric field. To bond the stress cone to the cable, a thin innermost layer of LDPE is necessary as bonding the cable directly to UHMWPE requires higher temperatures than the cable can withstand. Ribs are machined into the insulating section of the stress cone near the cable termination to reduce the likelihood of a discharge across the surface of the insulator. Since the stress cone will be attached to the HV cable and needs to pass through the HV cable conduit, the outer diameter and length of the stress cone were kept small, with the final dimensions to be determined after HV testing.

The stress cone components were machined (see Fig.~\ref{fig:stress_cone}B) and then bonded together in a copper mold to form a billet at \SI{200}{\celsius} for \SI{30}{\min}. Dimensions of the UHMWPE billet change through the baking process and require spring backed rams in the mold to restrict flow and maintain pressure, especially for the LDPE core. The billet tends to elongate axially and shrink on the diameter, which needs to be taken into account for the initial machined dimensions. Reducing air gaps between the machined pieces when bonding the UHMWPE billet is important to reduce void spaces in the final parts. After bonding, the billet is machined again to add the rib features and core out the center to fit precisely over the HV cable (with a tolerance of \SI{0.05}{in} (\SI{0.13}{\milli\meter})). 

Finally, the stress cone was bonded onto the high voltage cable with the outer ground (layer C) stripped back. A range of temperatures and dwell times were tested in order to achieve full penetration bonds (determined by visual inspection of sectioned prototypes) that withstood cryogenic testing. Ramping the oven temperature slowly is important as rapid ramping can cause the cable to lose shape and possibly fuse the different layers of the cable together, affecting the resistance and continuity of the cable. In our setup, baking the HV cable and stress cone at \SI{114}{\degreeCelsius} for \SI{90}{\min}, using the same ramp rates as the annealing process, produced bonds that withstood rapid cooling to liquid nitrogen temperatures and maintained the electrical properties of the cable.

After connection of the stress cone is complete, the insulating layer B at the end of the cable is trimmed to expose the conductive layer A for installation of the conductor termination. The conductor termination is composed of a phosphor bronze ``pill'' that mates to the HV connection near the base of the TPC. Phosphor bronze is used to avoid galling between the pill and the copper sphere. The pill is secured to the cable using screws that are sharpened to hold the components together as the cable cools. The cable conductor (layer A) is connected to the pill using a spring and copper crimp ferrule. A 6-point crimp tool is used to secure the ferrule on the central conductor while the spring seats around the ferrule and is compressed against the pill to make the electrical connection (see Fig.~\ref{fig:stress_cone}C).

\subsection{Angled connection}
Several designs were considered for the connection of the cable to the cathode. A fixed connection is more reliable but the presence of a \SI{\sim10}{\meter} cable connected to the bottom of the TPC vessel complicates the assembly and installation process and requires the welding of the copper conduit sections between the TPC and the top of the water tank with the HV cable inside. Here we present a design where the cable connection is removable and can be inserted after the TPC vessel and conduit are assembled and installed. 

For the connection we used a sphere-in-sphere geometry, which is a standard design in the gas-insulated switchgear (GIS) industry for coaxial connections~\cite{kuffel2000high}. Spheres are used as they minimize the electric field in the dielectric (in our case, liquid xenon) thereby eliminating the need for additional solid insulators to displace xenon from high field regions. It also allows for a wide range of angled connections with minimal change to the design. For example, changing from a 90\degree\ to 180\degree\ connection would allow for assembly of the HV connection with the cable fixed within the inner sphere. 

For an ideal spherical geometry (ignoring the necessary connection to the inner sphere) and a given potential difference $V$ between the inner and outer spheres, with radii $R_i$ and $R_o$ respectively, the maximum electric field can be analytically calculated to be 
\begin{align}
E_{max} = \frac{V}{R_i(1-R_i/R_o)}
\end{align}
%A plot of the maximum electric field as a function of the sphere diameters is shown for the nEXO cathode potential of \SI{50}{\kilo\volt} in Figure~\ref{fig:sphere_in_sphere}. 
For a given maximum design electric field, the size of the outer sphere is minimized for the ratio of $R_o = 2R_i$. 
%(shown by dotted line in Figure~\ref{fig:sphere_in_sphere}). 

\begin{figure*}
    \centering    \includegraphics[width=0.8\linewidth]{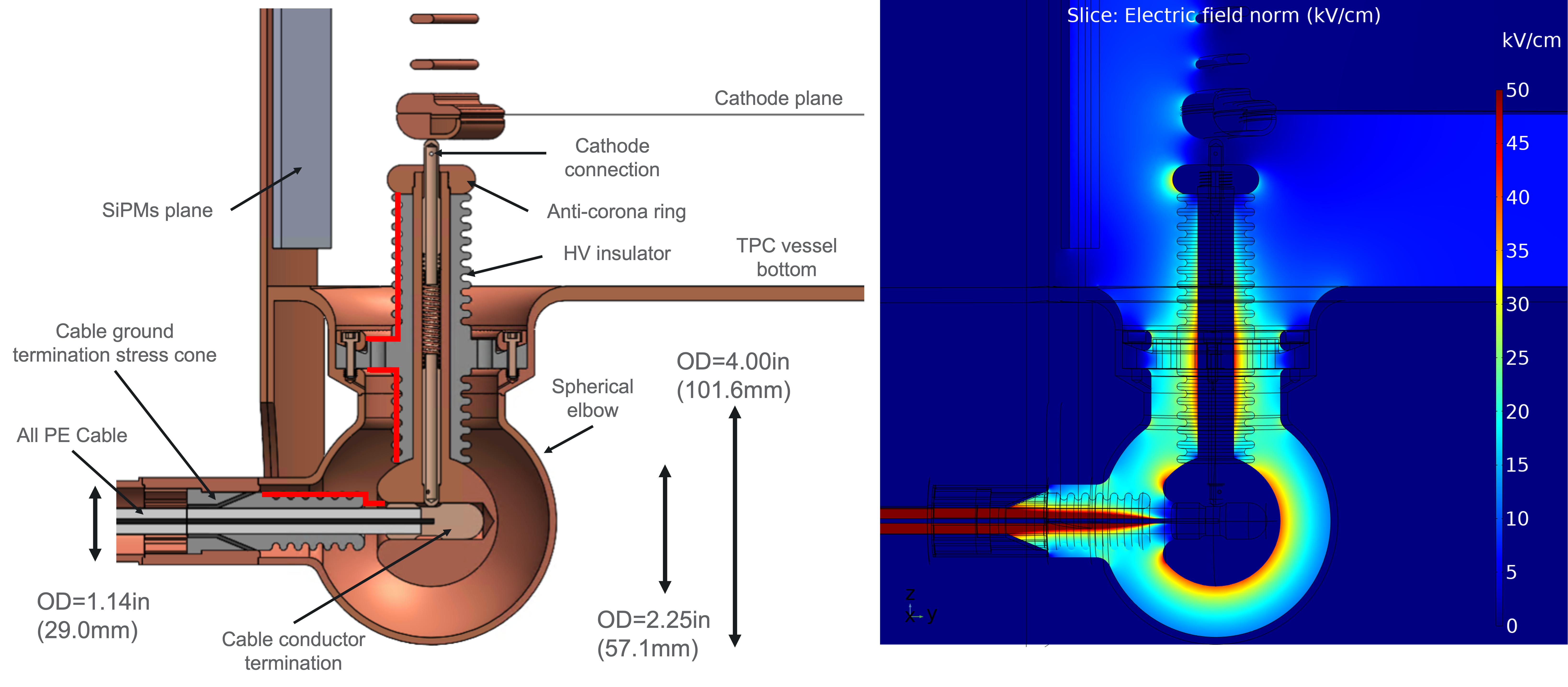}
    \caption{Left: Model of the HV connection of the cable to the cathode. Creepage paths from high voltage to ground are highlighted in red. Right: 3D electrostatic simulation showing the electric field strength in this region.}
    \label{fig:hv_connection}
\end{figure*}

Fig.~\ref{fig:hv_connection} shows a detailed model of the connection between the HV cable and the cathode, where we have attempted to incorporate all the best-practice design considerations discussed in Sec.~\ref{sec:hvdesign}. Based on a detailed 3D electrostatic FEA model in COMSOL\textsuperscript{\textregistered}~\cite{comsol2025} (Fig.~\ref{fig:hv_connection} right), the diameter of the inner and outer sphere were set at \SI{2.25}{in} (\SI{57.1}{\milli\meter}) and \SI{4}{in} (\SI{101.6}{\milli\meter}) respectively to keep the maximum electric fields in the xenon below the target \SI{50}{\kilo\volt\per\centi\meter}. Departure from the analytically ideal ratio is necessitated by the connections to the inner sphere. 

The potentially high-field triple point at the edge of the conductive pill is recessed within the inner sphere and all potential creepage paths (highlighted in Fig.~\ref{fig:hv_connection}) have ribs added with the minimum length being \SI{\sim 90}{\milli\meter} (from the stress cone ground to the inner sphere). The creepage lengths can be adjusted relatively easily in this design and the final values will be determined by testing in liquid xenon. Additionally, all edges have been rounded to reduce surface field enhancements and unavoidable sharp features (screw heads, threads, springs, etc.) have been shielded within equipotential conducting material.  

The connection between the inner sphere and the cathode is made through a vertical spring-loaded pin that contacts the underside of the cathode rim. The spring loaded design accommodates thermal contraction effects and possible seismic motion, ensuring continuous electrical connection to the cathode. The pin sits inside a HDPE insulator that is mechanically bolted to the copper TPC vessel and supports the inner sphere. To reduce the electric field at the termination of the insulator just below the cathode, we use a larger toroidal ``anti-corona'' ring. As for the other parts, all potential surface discharge paths have ribs added with creepage lengths of \SI{\sim 100}{\milli\meter} (inner sphere to TPC vessel) and \SI{\sim 145}{\milli\meter} (TPC vessel to anti-corona ring).
% all creepage lengths: stresscone->sphere = 88.21mm, sphere->plunger = 97.24mm, and plunger->cathode = 144.34mm

\begin{figure}
    \centering
    \includegraphics[width=\linewidth]{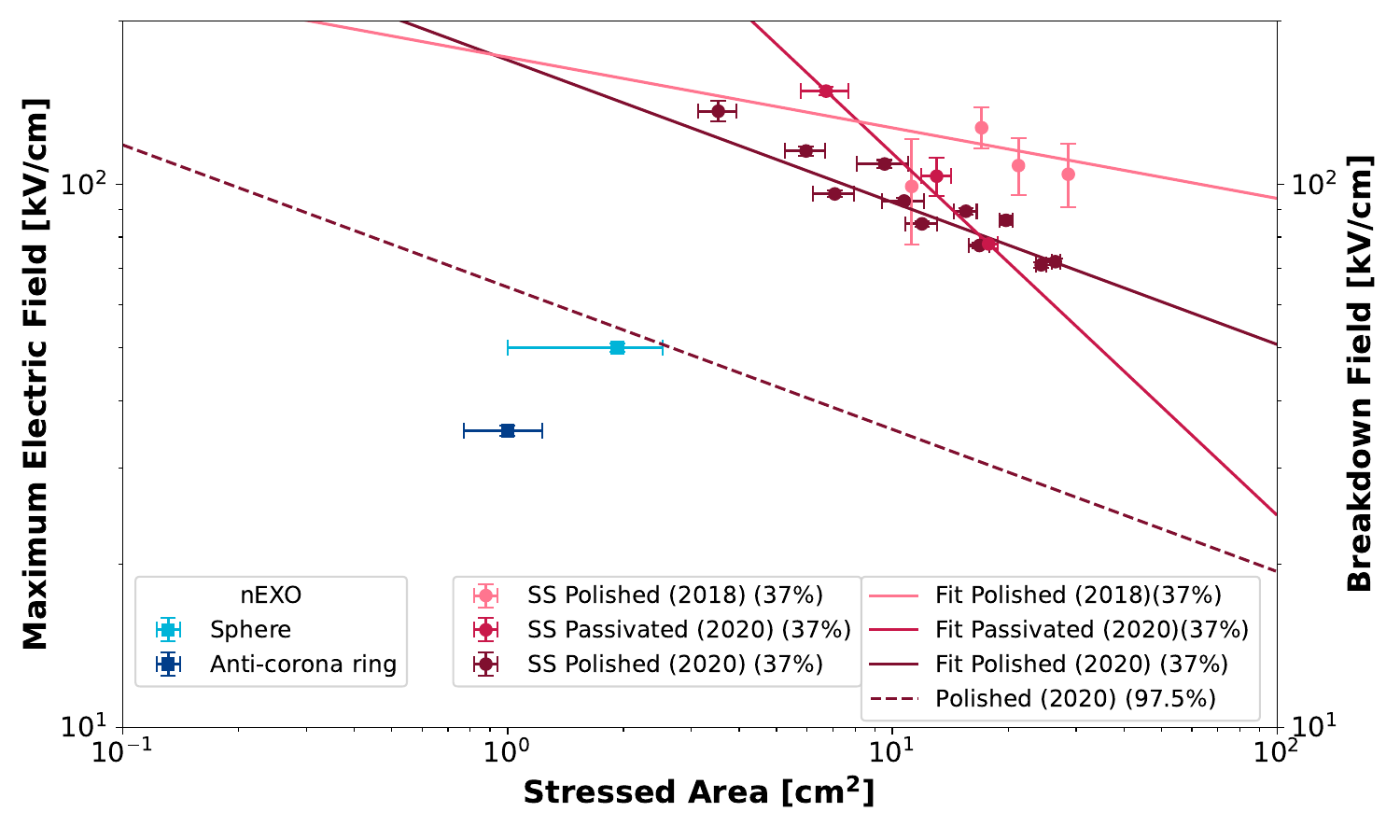}
    \caption{Plot of the highest surface electric fields and the corresponding stressed area of the inner sphere and anti-corona ring in the HV connection design (shades of blue). For comparison, we show the measured breakdown electric fields (at \SI{37}{\%} survival probability) and corresponding stressed areas for stainless steel (SS) Rogowski electrodes in the XeBrA experiment (shades of red), power law fits to the data (solid lines)~\cite{watson2023study}, and our extrapolation to \SI{97.5}{\%} survival probability for the 2020 polished SS data (dashed line). See text for details.}
    \label{fig:stressarea}
\end{figure}

The stressed areas of the highest field regions in liquid xenon, i.e., around the inner sphere and the ring at the cathode connection, were evaluated through the 3D electrostatic model. Using a fine mesh, the maximum electric field at the surface of the electrode in contact with the xenon was determined and then the total surface area with an electric field more than \SI{90}{\%} of that maximum value was calculated. Fig.~\ref{fig:stressarea} shows the maximum electric field and the corresponding stressed area values (in blue).

To our knowledge there have been no studies of breakdown voltage versus stressed area for copper electrodes in liquid xenon or other noble liquids. Nevertheless as a point of reference for our design (prior to conducting our own testing), we show the results from the XeBra experiment which studied high voltage breakdowns between two stainless steel Rogowski electrodes immersed in liquid xenon~\cite{tvrznikova2019direct, watson2023study}. The simulated electric field at breakdown was fitted with a 2- or 3-parameter Weibull distribution and the breakdown field corresponding to a \SI{37}{\%} survival probability was plotted versus the simulated stressed area, as shown in Fig.~\ref{fig:stressarea}. The corresponding power law fits for the area scaling factors from Ref.~\cite{watson2023study} are also shown. Assuming a Weibull distribution, one can also calculate the breakdown fields at higher survival probabilities, using the known relationship between the quantiles. The breakdown electric field $E_p$ for survival probability at the $p$ percentile follows the same scaling with stressed area $A$
\begin{align}
E_{p}(A) \propto \left(\ln \frac{100}{p}\right)^{1/k} A^{-1/k}
\label{eq:powerlaw}
\end{align}
where $k>0$ is the fitted shape parameter of the Weibull distribution.
As can be seen from Fig.~\ref{fig:stressarea} the peak electric field at both the copper sphere and anti-corona ring are beyond the \SI{97.5}{\%} survival probability curve (dashed line) for the polished stainless steel electrodes ($k=\num{3.82}$ from the 2020 data~\cite{watson2023study}) in liquid xenon. The distribution of electric field values at breakdown for the specific material and surface finish of these components will have to be experimentally verified.
% stress areas:
% hv plunger: a=1.00+-0.23cm2, Emax= 35.2+-0.8kV/cm
% sphere: 1.93-0.93+0.60cm2, Emax= 50.0+-0.9kV/cm
% cathode: 4.9-2.9+58.3cm2, Emax= 26.0+-0.5kV/cm
% bot. ring: 38.7+-3.7cm2, Emax= 21.3+-0.1kV/cm

\subsection{Test of stress cone and sphere-in-sphere}
Given the previous difficulties that noble liquid experiments have had with high voltage, it is imperative that HV designs are tested before being used. As described in Sec.~\ref{sec:hvdesign}, the stability of high voltage components depends critically on the electric field, stressed area, materials, and geometry of the conductors and insulators. It is therefore important to test the HV components at full-scale and in as close to final configuration as possible.

\begin{figure*}
    \centering    \includegraphics[width=0.8\linewidth]{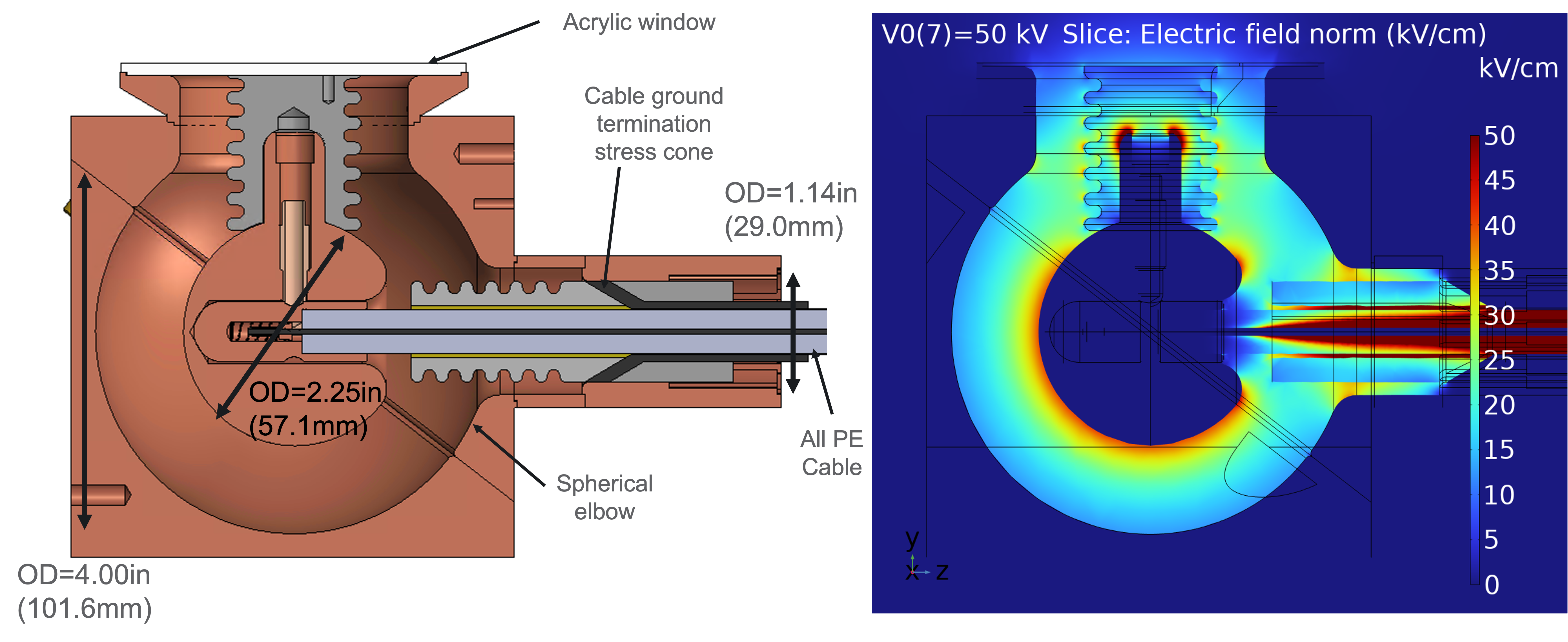}
    \caption{Left: Model of the full-scale HV connection prototype used for testing. Right: 3D electrostatic simulation showing the electric field strength.}
    \label{fig:hv_mockup}
\end{figure*}

A full-scale prototype of the HV connection, from the cable ground termination to the mechanical mounting of the insulator, see Fig.~\ref{fig:hv_mockup} left, was fabricated for testing. The prototype uses all the same materials (other than an acrylic window) as intended for the final component, and has the same stressed areas and creepage lengths due to the full-scale geometry. The only difference between the prototype and the nEXO design is that the vertical insulator that connects to the inner sphere is mounted to an acrylic window, where the insulator would normally mount to the TPC vessel. The window allows for monitoring of discharges between the inner and outer spheres.  To simplify the fabrication, the inner surface of the outer sphere is machined from a copper cube that is cut diagonally in two and then bolted together. A detailed 3D simulation of the electrostatic fields in the prototype setup was made to ensure the fields matched those in the nEXO detector design (see Fig.~\ref{fig:hv_mockup} right). Simulations of the electric field at the joint between the two halves of the cube showed that the field enhancement was minor compared to the electric fields on the surface of the inner sphere.
% creepage length is 82mm, ribs are 1/8in diameter

%\todo{mention other materials and surface finish}

As noted in Sec.~\ref{sec:hvdesign} the electrode surface finish can impact the stability at high voltage, therefore the regions of the prototype HV connection exposed to the highest electric fields (the inner surface of the external sphere and outer surface of the inner sphere) were polished after machining. The following polishing steps were applied: 1) hand polish with Simichrome~\cite{SimichromePol}; 2) wipe with AC500~\cite{AC500}; 3) sonicate for \SI{30}{\min} with Micro90~\cite{Micro90} at \SI{2}{\%} concentration; 4) rinse with DI water; and 5) surface wipe with ethanol. While the parts were thoroughly rinsed after polishing, radiopurity was not a concern for this test and no measurements were made to look for contamination of the copper surfaces. For parts that will be used in a low-background detector, a thorough evaluation of the cleaning procedure must be carried out and the procedure modified and components retested as needed (e.g. use of radiopure reagents and chemical etching to remove surface contaminants). The insulator parts were sonicated with ethanol after machining. 

Preliminary high voltage tests of the prototype HV connection were performed in air to verify some aspects of the design. Air was chosen for both simplicity and to accelerate the testing process (cheaper and faster turnaround than a cryogenic system using xenon). In a uniform field configuration with negligible influence of space charge, the breakdown voltage of gases through the Townsend mechanism is given by Paschen's Law. At atmospheric pressure and for electrode spacing between \SIrange{1}{10}{\cm}, Paschen's Law  gives the commonly cited electric field breakdown strength in air of roughly \SI{30}{\kilo\volt\per\centi\meter}~\cite{kuchler2017high}. While the electric field between the spheres in the HV connection (radial separation \SI{\sim 2.2}{\cm}) is not uniform, based on previous measurements of discharges in sphere-sphere air gaps in the literature we expect the maximum field at breakdown\footnote{The breakdown voltage does depend on air density and humidity but given the typical variations in our lab, the change is expected to be less than \qty{10}{\percent} \cite{kuffel2000high}.} to only be slightly larger than \SI{30}{\kilo\volt\per\centi\meter}~\cite{riba2024studying}. According to the field simulations of the prototype, this corresponds to an applied voltage of roughly \SI{-31}{\kilo\volt} on the inner sphere, with the outer sphere at ground. Note that given our weakly non-uniform field, we do not expect to see pre-discharges or corona that do not result in full breakdown~\cite{kuchler2017high}.

\begin{figure}
    \centering
\includegraphics[width=1\linewidth]{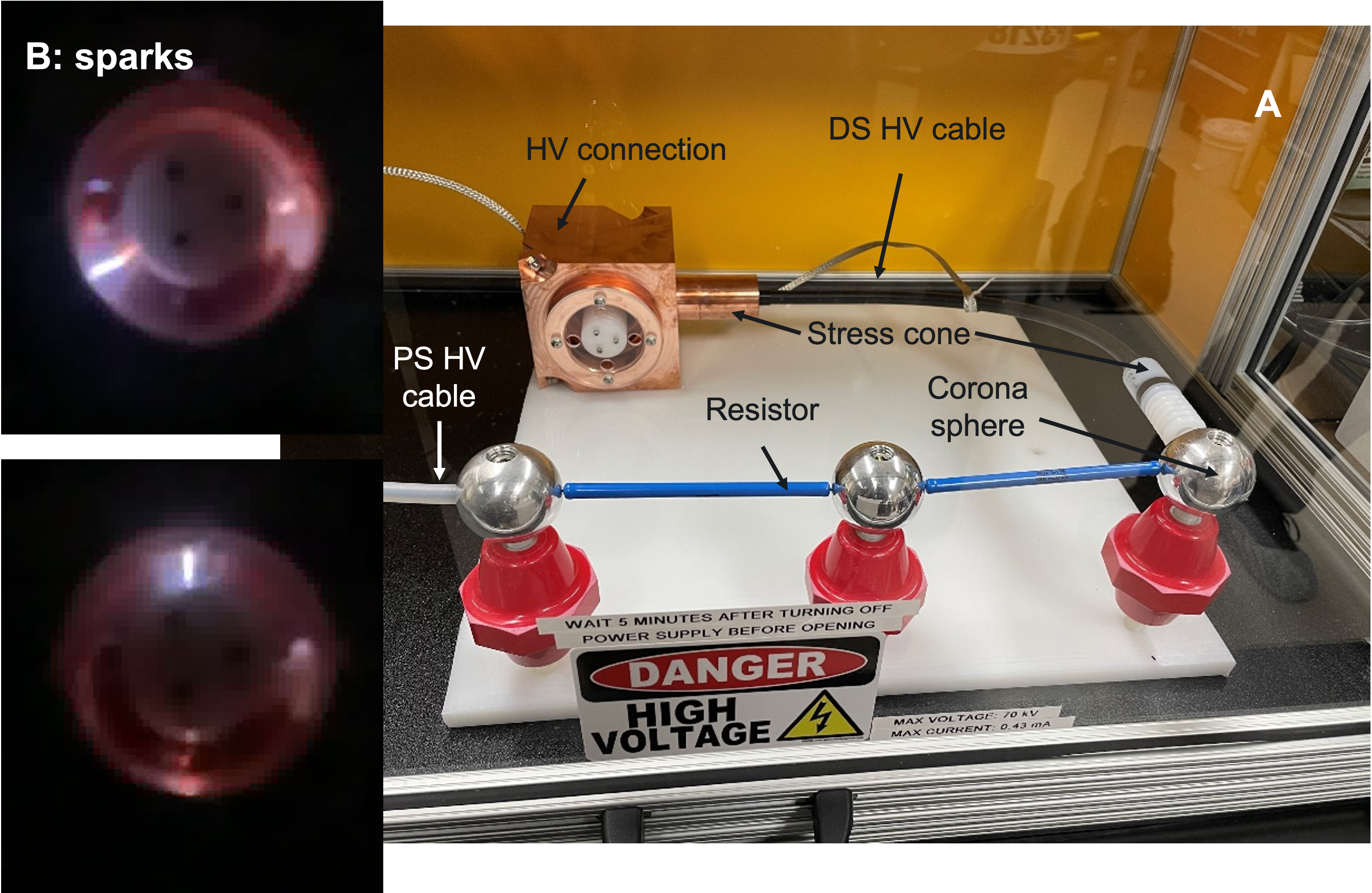}
    \caption{A: Picture of the components and setup for testing the HV connection (see Fig.~\ref{fig:hv_mockup}) in air. B: Examples of visible discharges seen between the inner and outer spheres.}
    \label{fig:air_test}
\end{figure}

Fig.~\ref{fig:air_test} shows the setup used for the air test. A \SI{-70}{\kilo\volt} SL70N30 Spellman power supply is connected to an interlocked HV enclosure via a standard HV cable with a metal core conductor. The standard HV cable is then connected to the HV connection prototype via a series of resistors (\SI{2}{\giga\ohm} total) and a short section of the DS-2353 PE cable featuring stress cones at both ends. The resistors are present to limit the current and avoid damage to electrodes during a discharge. The connections between the standard HV cable, the resistors, and the all-PE cable are made via corona spheres (\SI{47}{\milli\meter} OD) to avoid high fields at the connection points.

Two sets of tests were performed. In the initial test we were only aiming to verify that all the connections and components were correctly assembled and that breakdown only occurs at the expected voltage range. The high voltage was initially ramped up in increments of \SI{2.5}{kV/min} to \SI{-35}{\kilo\volt} and then more slowly at \SI{250}{V/min} up to \SI{-46}{\kilo\volt}. At \SI{-42.5}{\kilo\volt} visible and audible sparks appeared. The current output of the power supply was digitized and recorded for offline analysis. The charge drawn during each spark ranged from \SIrange{400}{2100}{\nano\coulomb}, with a typical charge of \SI{1900}{\nano\coulomb}. This is roughly consistent with the analytically calculated \SI{1260}{\nano\coulomb} of total charge stored in the sphere-in-sphere section modeled as a spherical capacitor (\SI{310}{\nano\coulomb}) and the short section of DS-2353 cable before the resistors (\SI{950}{\nano\coulomb}).

During the second test, the voltage was ramped up in \SI{5}{\kilo\volt} increment steps. After each step the voltage was maintained for \SIrange{3}{7}{\min} before proceeding to the next voltage setting. A few visible discharges with associated current draw started to occur at \SI{-25}{\kilo\volt} followed by a stable glow discharge (constant visible glow with a current draw of \SI{\sim30}{\micro\ampere}) at \SI{-30}{\kilo\volt} which faded and then disappeared when the voltage was lowered down to \SI{-20}{\kilo\volt}. On the following ramp up, visible discharges only occurred at or above \SI{-35}{\kilo\volt}. 

%\SI{35}{kV} in the course of 50 minutes,

In both tests, breakdowns were only observed in the air gap between the two spheres and the initiation points of these discharges were distributed over the surface of the spheres, indicating that there were no surface features with significantly enhanced local electric field. Though there was some variation, the voltage at breakdown is also roughly consistent with the expectation from Paschen's Law at atmospheric pressure. The results of these tests confirm that, as per design, the region with the highest fields in the insulating medium is between the spheres (and not, for example, at the cable ground termination) and the magnitude of this field roughly matches our expectation from electric field simulations. 

The final validation of the design of this HV connection needs to be performed in liquid xenon since the discharge mechanisms in liquid can be different from gas~\cite{sun2016formation, lesaint2016prebreakdown}. The goal of such tests will be to operate the full-scale components at high voltage in a liquid xenon environment that is as close to the final nEXO operational parameters (temperature, pressure, purity) as possible, characterize the voltage and location at which breakdown occurs, and look for precursors to breakdown with the glitch detector that can be used as indicators of instability in the final experiment. Once the design is finalized, long-term stability testing should be performed.

\section{Summary}
We have developed a design for the HV delivery system of the nEXO liquid xenon time projection chamber. The concept is based on HV design guidelines drawn from previous experimental studies of high voltage behavior in noble liquids, including studies carried out for EXO-200 experiment, as well as general design principles from the field of HV engineering. In addition to high voltage constraints, the design concept presented here meets the stringent radiopurity requirements set by nEXO's target sensitivity to neutrinoless double beta decay. While the complete integrated concept has yet to be fully tested in liquid xenon, several individual components have been tested and adopted by other experimental collaborations. We hope that these studies and the curation of references on high voltage design considerations will be useful for the design of HV components for other noble liquid TPCs.

\section{Acknowledgments}
This work was supported in part by Laboratory Directed Research and Development (LDRD) programs at Brookhaven National Laboratory (BNL), Lawrence Livermore National Laboratory (LLNL), Pacific Northwest National Laboratory (PNNL), and SLAC National Accelerator Laboratory. The authors gratefully acknowledge support for nEXO from the Office of Nuclear Physics within DOE’s Office of Science under grants/contracts DE-AC02-76SF00515, DE-FG02-01ER41166, DE-SC002305, DE-FG02-93ER40789, DE-SC0021388, DE-SC0012704, DE-AC52-07NA27344, DE‐SC0017970, DE-AC05-76RL01830, DE-SC0012654, DE-SC0021383, DE-SC0014517, DE-SC0024666, DE-SC0020509, DE-SC0024677 and support by the US National Science Foundation grants NSF PHY-2111213 and NSF PHY-2011948; from NSERC SAPPJ-2022-00021, CFI 39881, FRQNT 2019-NC-255821, and the CFREF Arthur B. McDonald Canadian Astroparticle Physics Research Institute in Canada; from IBS-R016-D1 in South Korea; and from CAS in China. 

\bibliographystyle{elsarticle-num-names}
\bibliography{bibliography.bib}

\end{document}